\documentclass[3p,times]{elsarticle}
\usepackage{ecrcpub}

\volume{952}

\firstpage{114934}

\journalname{Nuclear Physics B}

\runauth{T. Gutsche, V.E. Lyubovitskij, I. Schmidt}

\jid{npb}

\jnltitlelogo{Nuclear Physics B}

\usepackage{amssymb,amsmath,amsfonts}
\usepackage{epsfig,graphicx,tabularx}
\usepackage{color}
\usepackage[figuresright]{rotating}

\newcommand{\seq}{\begin{subequations}}
\newcommand{\sen}{\end{subequations}}
\newcommand{\eq}{\begin{eqnarray}}
\newcommand{\en}{\end{eqnarray}}
\newcommand{\ra}{\rangle}
\newcommand{\la}{\langle}

\begin{document}

\begin{frontmatter}

\dochead{}

\title{Electromagnetic properties of the nucleon and the Roper resonance\\       
in soft-wall AdS/QCD at finite temperature}

\author[label1]{Thomas Gutsche} 
\ead{thomas.gutsche@uni-tuebingen.de}
\author[label1,label2,label3]{Valery E. Lyubovitskij\corref{cor}}
\ead{valeri.lyubovitskij@uni-tuebingen.de} 
\author[label2]{Ivan Schmidt} 
\ead{ivan.schmidt@usm.cl} 
\address[label1]{Institut f\"ur Theoretische Physik,Universit\"at T\"ubingen,
Kepler Center for Astro and Particle Physics, \\
Auf der Morgenstelle 14, D-72076 T\"ubingen, Germany} 
\address[label2]{Departamento de F\'\i sica y Centro Cient\'\i fico
Tecnol\'ogico de Valpara\'\i so-CCTVal, \\ Universidad T\'ecnica
Federico Santa Mar\'\i a, Casilla 110-V, Valpara\'\i so, Chile} 
\address[label3]{Department of Physics, Tomsk State University,
634050 Tomsk, Russia}
\cortext[cor]{Corresponding author}

\begin{abstract}

We present a study of the nucleon electromagnetic form factors
and of the Roper-nucleon transition
at finite small temperature $T$, using an extended version
of a soft-wall AdS/QCD approach developed by us previously.
In the action we introduce the effective potential,
which has quadratic dependence on the holographic coordinate $z$
and depends on both the gluon and quark condensates.
Choosing the AdS geometry we restrict ourselves to the
AdS Poincar\'e metric, because the contribution of the
AdS-Schwarzschild geometry starts at next-to-leading order
${\cal O}(T^4)$. Hence, one can neglect the  temperature
dependence of the AdS geometry at small $T$.
This is consistent with the Hawking-Page phase transition
at a critical temperature representing the transition between 
thermal AdS/QCD and AdS-Schwarzschild geometry.
In the small temperature regime we base our analysis on the
temperature dependence of the effective potential,
which starts at order ${\cal O}(T^2)$,
due to the leading contribution from the quark condensate.
As applications we present the analysis of properties
of the nucleon and Roper resonance (masses, form factors,
and helicity amplitudes) at low temperatures.

\end{abstract}


\end{frontmatter}

\section{Introduction}

The study of baryons in the soft-wall AdS/QCD approach was first laid out  
in Refs.~\cite{Kirsch:2006he,Abidin:2009hr}, where an effective action was proposed, 
describing the nucleon confining dynamics and the minimal (nonminimal)
couplings to the electromagnetic field. Based on this action   
the electromagnetic form factors of the nucleon were calculated, and 
later, in Refs.~\cite{Vega:2010ns}-\cite{Gutsche:2019pls}, 
the soft-wall AdS/QCD approach in the baryon sector was modified/improved 
in different directions. It was applied to the calculation of 
generalized parton distributions of the nucleon~\cite{Vega:2010ns}, 
extended to account for higher Fock states 
in the nucleon and to additional couplings with the electromagnetic 
field~\cite{Gutsche:2012bp}. This last extension leads to consistency 
with QCD constituent counting rules~\cite{Brodsky:1973kr} 
for the power scaling of hadronic form factors at large values of the 
Euclidean momentum transfer squared. In Ref.~\cite{Gutsche:2011vb} 
the soft-wall AdS/QCD approach was further developed to describe baryons 
with adjustable quantum numbers $n$, $J$, $L$, and $S$.  
In another development, in Refs.~\cite{Brodsky:2014yha,Sufian:2016hwn},  
the nucleon properties were analyzed using a Hamiltonian 
formalism. In Ref.~\cite{FolcoCapossoli:2019imm} a version 
of the soft-wall AdS/QCD approach was proposed, where a modified 
warp factor in the metric tensor is present. 
Note that in Ref.~\cite{Gutsche:2011vb} we proved that any modification 
of the warp factor in the metric tensor can be compensated by an appropriate 
choice of the holographic potential. The form of such potentials for 
AdS fields with different spins were also derived analytically. 

The soft-wall AdS/QCD formalism is also useful for constraining other approaches. 
In particular, in Refs.~\cite{Gutsche:2013zia,Gutsche:2016gcd}  
we developed a light-front quark-diquark approach to nucleon structure 
describing nucleon parton distributions and form factors from a unified 
point of view. In particular, in Ref.~\cite{Gutsche:2016gcd} we derived 
nucleon light-front wave functions, analytically matching the results of 
global fits to the quark parton distributions in the nucleon at the initial 
scale $\mu \sim 1$ GeV. We also showed that these distributions obey the correct
Dokshitzer-Gribov-Lipatov-Altarelli-Parisi evolution~\cite{DGLAP} up to 
high scales. Using these constraints for the nucleon wave functions, we got a reasonable
description of data of the nucleon electromagnetic form factors. We also made 
predictions for other nucleon quark distributions (transverse momentum, 
Wigner and Husimi distributions) from a unified point of view, using our 
light-front wave functions and expressing them in terms of the parton 
distributions $q_v(x)$ and $\delta q_v(x)$.

Nucleon resonances have also been discussed in the AdS soft-wall approach 
in Refs.~\cite{deTeramond:2011qp}-\cite{Gutsche:2017lyu}.     
In Ref.~\cite{deTeramond:2011qp} the Dirac form factor for the
electromagnetic nucleon-Roper transition was calculated in
light-front holographic QCD.
Later, in Ref.~\cite{Gutsche:2012wb},
a formalism for the study of all nucleon resonances in soft-wall
AdS/QCD has been proposed. As a first application, a detailed
description of Roper-nucleon transition properties
(form factors, helicity amplitudes and transition charge radii)
was performed. 
In Ref.~\cite{Gutsche:2017lyu} we presented an improved study of the 
electromagnetic form factors of the nucleon and of the Roper, 
using an extended version of the effective action of 
soft-wall AdS/QCD. At this point  additional 
non-minimal terms were included, which do not renormalize the charge 
and give an important contribution to the momentum dependence of 
the form factors and helicity amplitudes. 

The study of hadron properties at finite temperature gives a unique 
opportunity for the understanding of the evolution of the early Universe 
after the Big Bang. It also permits to investigate physical principles 
related to the formation of hadronic matter and its phase transitions 
(confinement and deconfinement, chiral symmetry restoration). 
From experiments, e.g., on quark-gluon plasma formation, we know that hadron 
properties change under the influence of in-medium effects (density and 
temperature). Therefore, the investigation of the temperature behavior 
of hadrons is a worthwhile task to pursue.
One of the most interesting hadronic characteristics 
to study is the temperature dependence of masses and form factors. 
In Refs.~\cite{Gutsche:2019blp,Gutsche:2019pls} 
we derived a soft-wall AdS-Schwarzschild approach at small temperatures, in which 
we included temperature dependence due to the propagation 
of AdS fields in the AdS-Schwarzchild metric and due to the explicit $T$ 
dependence of the dilaton scale parameter $\kappa$. We performed a temperature 
expansion of hadronic quantities in the form of an $T^2/(12 F^2)$ expansion 
dictated by QCD~\cite{Gasser:1986vb}-\cite{Toublan:1997rr}  
and noticed that the AdS-Schwarzschild thermal factor $f_T(z) = 1 - (\pi T z)^4$ 
gives a contribution at next-to-leading order, i.e. at 
order ${\cal O}(T^4)$, while the $T$ dependence of the dilaton 
shows up already at leading order ${\cal O}(T^2)$. 
Our approach was applied to the description of hadrons with integer 
and half-integer spin and adjustable number of constituents 
(mesons, baryons, tetraquarks, etc.) with analytical results for 
the temperature dependence of their masses and form factors. 
Note that originally the idea of a thermal dilaton --- dilaton field 
depending on temperature was proposed in Ref.~\cite{Vega:2018dgk}. 
Two thermal dilaton forms were studied resulting in 
melting temperatures for mesons close to 180 MeV.

In the present paper, using our findings of 
Refs.~\cite{Gutsche:2019blp,Gutsche:2019pls}, we present a more 
detailed discussion of the effective holographic potential 
providing spontaneous breaking of conformal and chiral symmetry 
and generating a temperature dependence. Using the 
fact that the holographic potential has leading ${\cal O}(T^2)$ 
we restrict to the use of the pure AdS Poincar\'e metric, which 
is more relevant in small temperature regime in comparison with 
the AdS-Schwarzchild geometry, which is more suitable at high $T$. 

The paper is organized as follows. 
In Sec.~II we briefly discuss our formalism. 
In Sec.~III we present the analytical calculation and the numerical analysis 
of electromagnetic form factors and helicity amplitudes of the nucleon 
and the Roper at finite temperature. 
Finally, Sec.~IV contains our summary and conclusions.

\section{Formalism}

In this section we briefly review our approach~\cite{Gutsche:2017lyu}. 
We start with the definition of the conformal Poincar\'e metric 
\eq
g_{MN} \, x^M x^N = \epsilon^a_M \, \epsilon^b_N \, 
\eta_{ab} \, x^M x^N = \frac{1}{z^2} \, (dx_\mu dx^\mu - dz^2)
\en 
where $\epsilon^a_M = \delta^a_M/z$ is the              
vielbein, and we define $g = |{\det}(g_{MN})| = 1/z^{10}$ as the magnitude 
of the determinant of $g_{MN}$. 

The soft-wall AdS/QCD action $S$ for the nucleon 
$N = (p,n)$ and the Roper ${\cal R} = ({\cal R}_p,{\cal R}_n)$ resonance, 
including photons, is constructed in terms of the dual spin-$1/2$ fermion and 
vector fields. They have constrained (confined) dynamics in AdS space 
due to the presence of a quadratic ($z^2$) holographic potential. 
One can derive such a potential in the action: (1) via an exponential prefactor 
containing the background field or (2) directly as an interaction potential. 
We proved in Ref.~\cite{Gutsche:2011vb} that both versions of the soft-wall AdS/QCD model 
are equivalent to each other. As was first shown in Ref.~\cite{Kirsch:2006he} 
and later confirmed in Ref.~\cite{Abidin:2009hr}, 
in the case of baryons the exponential prefactor can be simply removed after proper 
redefinition of the AdS fermion fields, which means that the direct inclusion of the 
holographic potential is essential. Such a potential was constructed for the first time in 
Refs.~\cite{Kirsch:2006he,Abidin:2009hr}:  
$U_F(z) = \varphi(z) = \kappa^2 z^2$, where $\varphi$ is the quadratic dilaton 
field and $\kappa$ is its size parameter. 
Keeping in mind that the most fundamental property of baryons 
is their mass. It can be expressed in terms of vacuum condensates 
using methods of QCD sum rules~\cite{Ioffe:1981kw,Thomas:2001kw}, 
where at leading order the nucleon mass is related to the quark 
condensate (so-called Ioffe formula)~\cite{Ioffe:1981kw} by:   
\eq\label{MN_Ioffe2}
M_N^3 \simeq 8 \, \pi^2 \, \Sigma \,, 
\en
where $\Sigma = |\la 0| \bar q q | 0 \ra|$ is the quark condensate. 
The Ioffe formula gives reasonable agreement with data. 
However, as stressed in Ref.~\cite{Gorsky:2013dda}, there is 
a puzzle, since this formula implies that a large contribution to baryon masses 
comes from the quark condensate. In the overview to this problem 
presented in Ref.~\cite{Gorsky:2013dda} it is mentioned 
that conceptually different theoretical approaches 
(lattice QCD~\cite{Glozman:2012fj}, chiral quark-soliton 
model~\cite{Diakonov:1987ty}, dilaton compensated model 
based on hidden local symmetry~\cite{Ma:2013ela}, 
holographic QCD~\cite{Gorsky:2015pra}) arrived 
at similar conclusions, which is that the main source of the baryon 
mass is due to the quark condensate independent term $m_0$  
plus a correction due to the quark condensate. 
In particular, in Ref.~\cite{Ma:2013ela} 
the following formula for the baryon mass $M_B$ was proposed: 
\eq 
M_B = m_0 + \Delta(\Sigma) \,,
\en 
where $\Delta(\Sigma)$ is the correction due to the 
quark condensate. 
In Ref.~\cite{Gorsky:2015pra} it was found that 
there are two different regimes of low and large values of 
the quark condensate $\Sigma = (N_c/2\pi^2) \, \sigma$. 
At low values of $\sigma$ the baryon mass is saturated by the 
condensate independent term $887$ MeV controlled by the confinement scale, 
while at large values of $\sigma$ it is defined by the scalar repulsion 
force and linearly depends on the quark condensate~\cite{Gorsky:2015pra}: 
\eq 
M_B = {\rm max}[887; -407 + 940 \, (z_m \sigma^{1/3})] \ {\rm MeV}\,, 
\en   
where $z_m = (323 \ {\rm MeV})^{-1}$ is the position of the infrared 
hard wall in the holographic approach that we are considering. 
Note that the physical explanation of the dominant term $m_0$ 
is different in various approaches: constituent quark mass, 
qluon condensate, skyrmion, etc. 

In our approach the baryon masses are proportional to the 
scale parameter $\kappa$ occurring in the $z^2$ dependent 
holographic potential $U_F(z) = \kappa^2 z^2$. 
This corresponds to the square of the 
linearized part of the Cornell potential, 
because the equation of motion (EOM) for the wave function with $U_F$ 
corresponds to the square of the EOM with the Cornell potential, 
as proven in Refs.~\cite{Trawinski:2014msa,Gutsche:2014oua}. 
Here $\kappa = 383$ MeV is the dilaton scale parameter fixed 
in the previous study. 
In order to get explicit expressions for the baryon masses in 
terms of quark condensates we parametrize the scale parameter as:  
\eq\label{kappa2_expansion} 
\kappa^2 = \kappa^2 \, (1-y) \,+\, \frac{\pi^2 \, \Sigma}{M_N} \, y\,,
\en 
where $y$ is the mixing parameter. The specific normalization 
of the condensate containing term is chosen for convenience. 
The squared nucleon mass in the soft-wall AdS/QCD model, 
for leading twist $\tau = 3$, is given by~\cite{Gutsche:2011vb}: 
\eq 
M_N^2 = 8 \, \kappa^2\,, 
\en   
then one gets: 
\eq 
M_N^2 = 8 \, \kappa^2 \, (1-y) \,+\,
\frac{8\, \pi^2\, \Sigma}{M_N} \, y\,. 
\en 
Next, using the Ioffe formula~(\ref{MN_Ioffe2}) one has 
\eq 
M_N^2 = 8 \, \kappa^2 \, (1-y) \,+\,  M_N^2 \, y\,. 
\en 
This last relation has a very clear physical meaning.  
Here and in Eq.~(\ref{kappa2_expansion}) the limits $y=0$ and 
$y=1$ correspond to the cases of vanishing and fully dominating 
quark condensates in the expression for the nucleon mass and $\kappa^2$.  
In previous papers we consider the case $y=1$. Here we vary the parameter $y$ 
and investigate the dependence on $y$ of the nucleon and Roper resonance 
properties, and we do not specify the physical content 
of the parameter $\kappa^2$, which 
might be a combination of gluon condensates of dimension 2 and 4. 
It is important that these gluon condensates have a suppressed 
temperature dependence at small $T$ in comparison with the quark 
condensate. In particular, the temperature dependence of the 
dimension-2 gluon condensate ${\cal O}_{A^2} = \la A_\mu^2 \ra$ has been 
analyzed in a formalism of local composite 
operators~\cite{Vercauteren:2010rk}, where tt was shown that no 
${\cal O}(T^4)$ temperature corrections arise for this condensate at small $T$. 
The study of the temperature dependence of the dimension-4 gluon condensate 
${\cal O}_{G^2} = \la {\rm tr}\,G_{\mu\nu}^2 \ra$ has been initiated by 
Leutwyler in Ref.~\cite{Leutwyler:1992cd} and later it was extensively 
studied, e.g., in $SU(2)$ and $SU(3)$ lattice gauge theories in 
Ref.~\cite{Miller:1997dn}. There it was shown that in both cases 
the temperature corrections start at order ${\cal O}(T^4)$, but 
are small numerically, making the condensate to be constant 
for small $T < T_c$, where $T_c = 290$ MeV for $SU(2)$ and 
$264$ MeV for $SU(3)$.  Therefore, the main source of the temperature 
dependence of the holographic potential in the low $T$-region 
comes from the quark condensate. The temperature dependence of the 
holographic potential will disappear when $\Sigma$ vanishes which corresponds 
to the restoration of spontaneously broken chiral symmetry. 
In QCD the quark condensate depends on temperature and vanishes at the critical 
temperature $T_c$, which signals the restoration of chiral symmetry.
Such temperature dependence of the quark condensate up to $T_c$, 
was calculated by Gasser, Leutwyler, and Gerber using chiral perturbation 
theory (ChPT)~\cite{Gasser:1986vb,Leutwyler:1987th} in the form 
of an $T^2/(12 F^2)$ expansion dictated by QCD at two~\cite{Gasser:1986vb} 
and three loops~\cite{Leutwyler:1987th}. In particular, the result 
for two-loop ChPT reads:      
\eq\label{SigmaT_2loops} 
\Sigma(T) &=& \Sigma \ \sigma(T)\,, \nonumber\\
\sigma(T) &=& 1 + \delta_{T_1} \frac{T^2}{12 F^2}
\,+\, \delta_{T_2} \biggl(\frac{T^2}{12 F^2}\biggr)^2
\,+\,{\cal O}(T^6) \,, \nonumber\\
\delta_{T_1}  &=&  - \frac{N_f^2-1}{N_f}     \,, \quad
\delta_{T_2} \,=\, - \frac{N_f^2-1}{2 N_f^2} \,, 
\en  
where $F$ is the pseudoscalar coupling constant in the chiral limit, 
and $N_f$ is the number of quark flavors. 
Note that the quark condensate has an entirely non-perturbative origin, while 
its $T$-dependence has been calculated in an
expansion in powers of $T^2/(12 F^2)$ using ChPT. 
The thermal quark condensate is normalized at zero temperature 
as $\Sigma(0) = \Sigma$, while it vanishes $\Sigma(T_c) = 0$ 
at the critical temperature $T_c$, i.e. at the temperature when 
the sponeously broken chiral symmetry is restored. 
Therefore, the temperature dependence of the holographic potential 
$U_F(z)$ can be fixed by the temperature dependence of the quark 
condensate, which has been evaluated using ChPT: 
\eq\label{U_F}
U_F(z,T) = \kappa^2(T) \, z^2\,, 
\en   
where $\kappa^2(T)$ is the temperature dependent scale parameter:  
\eq 
\kappa^2(T) = \kappa^2 \, (1-y) \,+\, \frac{\pi^2 \, \Sigma(T)}{M_N} \, y \,. 
\en             
Using the relations derived above one can rewrite $\kappa^2(T)$ in a more 
convenient form 
\eq 
\kappa^2(T) = \kappa^2 \bigg( 1 + y \, \Big[ \sigma(T) - 1 \Big] \biggr) \,. 
\en 

In our numerical analysis we present results for the quark condensate below 
the critical temperature $T_c$, 
which is fixed from the condition $\sigma(T_c) = 0$. 
Using Eq.~(\ref{SigmaT_2loops}) we have  
\eq 
\frac{T_c^2}{12 F^2} 
= N_f \, \biggl[
\sqrt{\frac{N_f^2+1}{N_f^2-1}} - 1 \biggr] \,.
\en
For the case in which the number of quark flavors is set to $N_f = 2$ 
one gets $T_c = 179$~MeV. Therefore, the study of the $T$-dependence of
baryon properties provides a good possibility to check the contribution of 
the quark condensate to their observables, e.g., masses and form factors.

Now let us specify the action relevant for the study of properties of 
the nucleon and Roper resonance. 
The action $S$ contains a free part $S_0$, describing the 
confined dynamics of AdS fields, and an interaction part $S_{\rm int}$, 
describing the interactions of fermions with the vector field    
\eq\label{actionS}
S   &=& S_0 + S_{\rm int}\,, \nonumber\\[3mm]
S_0 &=& \int d^4x dz \, \sqrt{g} \, e^{-\varphi(z,T)} \,
\biggl\{ {\cal L}_N(x,z,T) + {\cal L}_{\cal R}(x,z,T)
+ {\cal L}_V(x,z,T)
\biggr\} \,, \nonumber\\[3mm]
S_{\rm int} &=& \int d^4x dz \, \sqrt{g} \, e^{-\varphi(z,T)} \,
\biggl\{
{\cal L}_{VNN}(x,z) + {\cal L}_{V{\cal R}{\cal R}}(x,z,T)
+ {\cal L}_{V{\cal R}N}(x,z,T)
\biggr\} \,. 
\en
where ${\cal L}_N$, ${\cal L}_{\cal R}$, ${\cal L}_V$ and
${\cal L}_{VNN}$, ${\cal L}_{V{\cal RR}}$,
${\cal L}_{V{\cal R}N}$ are the free and
interaction Lagrangians,
respectively, which are written as 
\eq
{\cal L}_B(x,z,T) &=&  \sum\limits_{i=+,-; \,\tau} \, c_\tau^B \,
\bar\psi^B_{i,\tau}(x,z,T) \, \hat{\cal D}_i(z) \, \psi^B_{i,\tau}(x,z,T)
\,, \nonumber\\[3mm]
{\cal L}_V(x,z,T) &=& - \frac{1}{4} V_{MN}(x,z,T)V^{MN}(x,z,T)\,, 
\nonumber\\[3mm]
{\cal L}_{VBB}(x,z,T) &=& \sum\limits_{i=+,-; \tau} \, c_\tau^B \,
\bar\psi^B_{i, \tau}(x,z,T) \, \hat{\cal V}^B_i(x,z,T) \,
\psi^B_{i, \tau}(x,z,T)\,, \nonumber\\[3mm]
{\cal L}_{V{\cal R}N}(x,z,T) &=&
\sum\limits_{i=+,-; \,\tau} \, c_\tau^{{\cal R}N} \,
\bar\psi_{i,\tau}^{\cal R}(x,z,T) \, \hat{\cal V}^{{\cal R}N}_i(x,z,T) \,
\psi_{i,\tau}^N(x,z,T) \, + \, {\rm H.c.}\,,
\en
where $B = N, {\cal R}$ and
\eq
\hat{\cal D}_\pm(z,T) &=&  \frac{i}{2} \Gamma^M
\! \stackrel{\leftrightarrow}{\partial}_{_M} - \frac{i}{8}
\Gamma^M \omega_M^{ab} [\Gamma_a, \Gamma_b]
\, \mp \,  (\mu + U_F(z,T))\,, \nonumber\\[3mm]
\hat{\cal V}^H_\pm(x,z,T)  &=&  Q \, \Gamma^M  V_M(x,z,T) \, \pm \,
\frac{i}{4} \, \eta_V^H \,  [\Gamma^M, \Gamma^N] \, V_{MN}(x,z,T) 
\, \pm \, \frac{i}{4} \, \lambda_V^H \, z^2 \, 
[\Gamma^M, \Gamma^N] \, \partial^K\partial_KV_{MN}(x,z,T)  
\nonumber\\[3mm] 
&\pm& \, g_V^H \, \Gamma^M \, i\Gamma^z \, V_M(x,z,T)
+     \zeta_V^H \, z \, \Gamma^M  \, \partial^N V_{MN}(x,z,T)
\, \pm \, \xi_V^H \, z \, \Gamma^M \, i\Gamma^z \, \partial^N V_{MN}(x,z,T)
\,.
\en
Here $H = N, {\cal R}, {\cal R}N$, 
$\mu$ is the five-dimensional mass of the spin-$\frac{1}{2}$ AdS      
fermion, $\mu = 3/2 + L$, with $L$ being the orbital angular momentum;      
$U_F(z,T) = \varphi(z,T)$ is the dilaton potential; 
$Q = {\rm diag}(1,0)$ is the nucleon (Roper) charge matrix;  
$V_{MN} = \partial_M V_N - \partial_N V_M$ is                            
the stress tensor for the vector field; 
$\omega_M^{ab} = (\delta^a_M \delta^b_z - \delta^b_M \delta^a_z)/z$ is 
the spin connection term; and $\sigma^{MN} = [\Gamma^M, \Gamma^N]$ 
is the commutator of the Dirac matrices in AdS space, which are defined as  
$\Gamma^M = \epsilon^M_a \Gamma^a$ and 
$\Gamma^a = (\gamma^\mu, -i \gamma^5)$.   

The action~(\ref{actionS}) is constructed in terms of the 
5D AdS fermion fields $\psi^N_{\pm,\tau}(x,z,T)$, 
$\psi^{\cal R}_{\pm,\tau}(x,z,T)$, and the vector field $V_M(x,z,T)$. 
Fermion fields are duals to the left- and 
right-handed chiral doublets of nucleons and Roper resonance 
${\cal O}^L = (B_1^L, B_2^L)^T$ and ${\cal O}^R = (B_1^R, B_2^R)^T$, 
with $B_1 = p, {\cal R}_p$ and $B_2 = n, {\cal R}_n$. These fields  
are in the fundamental representations of the chiral $SU_L(2)$ 
and $SU_R(2)$ subgroups and are holographic analogues of the nucleon $N$ 
and Roper resonance ${\cal R}$, respectively. 

The 5D AdS fields $\psi^B_{\pm,\tau}(x,z,T)$ are products of the left/right 
4D spinor fields 
\eq\label{psi_expansion}
\psi^{L/R}_{n=0,1}(x,T) = \frac{1 \mp \gamma^5}{2} \, \psi_{n=0,1}(x,T)\,,
\en
with spin $1/2$ and bulk profiles 
\eq\label{psi_expansion1}
F^{L/R}_{\tau, n=0,1}(z,T) = z^2 \, f^{L/R}_{\tau, n=0,1}(z,T),
\en 
with twist $\tau$ depending on the holographic (scale) variable $z$: 
\eq\label{psi_expansion2}
\psi^N_{\pm,\tau}(x,z,T) &=& \frac{1}{\sqrt{2}} \,
\left[
      \pm \psi^{L}_0(x,T) \ F^{L/R}_{\tau, 0}(z,T)
+   \psi^{R}_0(x,T) \ F^{R/L}_{\tau, 0}(z,T)\right]\,, \nonumber\\
\psi^{\cal R}_{\pm,\tau}(x,z,T) &=& \frac{1}{\sqrt{2}} \,
\left[
    \pm  \psi^{L}_1(x,T) \ F^{L/R}_{\tau, 1}(z,T)
+   \psi^{R}_1(x,T) \ F^{R/L}_{\tau, 1}(z,T)\right]\,,
\en
where
\eq\label{fL_fR} 
f^L_{\tau, 0}(z,T) &=& \sqrt{\frac{2}{\Gamma(\tau)}} \, \kappa^{\tau}(T) \,
z^{\tau - 1/2} \, e^{-\kappa^2(T) z^2/2}\,, \nonumber\\
f^R_{\tau, 0}(z,T)  &=& \sqrt{\frac{2}{\Gamma(\tau-1)}} \, \kappa^{\tau-1}(T) \,
z^{\tau - 3/2} \, e^{-\kappa^2(T) z^2/2}\,, \nonumber\\
f^L_{\tau, 1}(z,T)  &=& \sqrt{\frac{2}{\Gamma(\tau+1)}} \, \kappa^{\tau}(T) \,
z^{\tau - 1/2} \, (\tau  - \kappa^2(T) z^2) \, e^{-\kappa^2(T) z^2/2}\,, 
\nonumber\\
f^R_{\tau, 1}(z,T)  &=& \sqrt{\frac{2}{\Gamma(\tau)}} \, \kappa^{\tau-1}(T) \,
z^{\tau - 3/2} \, (\tau - 1 - \kappa^2(T) z^2) \, e^{-\kappa^2(T) z^2/2}\,.
\en
The nucleon is identified as the ground state with $n=0$
and the Roper resonance as the first radially excited state with $n=1$. 
In the case of the vector field we work in the axial gauge
$V_z = 0$ and perform a Fourier transformation of 
the vector field $V_\mu(x,z,T)$ with respect to the Minkowski coordinate
\eq\label{V_Fourier}
V_\mu(x,z,T) = \int \frac{d^4q}{(2\pi)^4} e^{iqx} V_\mu(q) V(q,z,T)\,. 
\en
We derive an EOM for the vector bulk-to-boundary propagator $V(q,z)$,
dual to the $q^2$-dependent electromagnetic current
\eq
\partial_z \biggl( \frac{e^{-\varphi(z,T)}}{z} \,
\partial_z V(q,z,T)\biggr) + q^2 \frac{e^{-\varphi(z,T)}}{z} \, 
V(q,z,T) = 0 \,. 
\en 
The solution of this equation in terms of the
gamma $\Gamma(n)$ and Tricomi $U(a,b,z)$ functions reads
\eq
\label{VInt_q}
V(q,z,T) = \Gamma\Big(1 - \frac{q^2}{4\kappa^2(T)}\Big)
\, U\Big(-\frac{q^2}{4\kappa^2(T)},0,\kappa^2(T) z^2\Big) \,.
\en
In the Euclidean region ($Q^2 = - q^2 > 0$)
it is convenient to use the integral
representation for $V(Q,z,T)$~\cite{Grigoryan:2007my}
\eq
\label{VInt}
V(Q,z,T) = \kappa^2(T) z^2 \int_0^1 \frac{dx}{(1-x)^2}
\, x^{a_T} \,
e^{- \kappa^2(T) z^2 \frac{x}{1-x} }\,,
\en
where $x$ is the light-cone momentum fraction and
$a_T = Q^2/(4 \kappa^2(T))$.

The set of parameters 
$c_\tau^{N}$, $c_\tau^{{\cal R}}$, and $c_\tau^{{\cal R}N}$ 
induce mixing of the contribution of AdS 
fields with different twist dimensions. 
In Refs.~\cite{Gutsche:2012bp,Gutsche:2012wb} we
showed that the parameters $c_\tau^B$  are constrained 
by the condition $\sum_\tau \, c_\tau^B = 1$, in order to get
the correct normalization of the kinetic term
$\bar\psi_n(x)i\!\not\!\partial\psi_n(x)$
of the four-dimensional spinor field. This condition is also
consistent with electromagnetic gauge invariance. 

The couplings $G_V^H = {\rm diag}(G_V^{H_1},G_V^{H_2})$,  
where $G = \eta, \lambda, g, \zeta, \xi$, 
$\ H_1 = p, {\cal R}_p, {\cal R}_pp$ and 
$\ H_2 = n, {\cal R}_n, {\cal R}_nn$ are fixed from
the magnetic moments, slopes, and form factors
of both the nucleon and Roper,
while the couplings $c_\tau^{{\cal R}N}$ are fixed from
the normalization of the Roper-nucleon helicity amplitudes.
The terms proportional to the couplings $\lambda_V^H$, $\zeta_V^H$, 
and $\xi_V^H$ express novel nonminimal couplings of the fermions 
with the vector field. 
These couplings do not renormalize the charge and do not change 
the corresponding form factor normalizations, but give 
an important contribution to the momentum dependence of 
the form factors and helicity amplitudes. 

The nucleon and Roper masses are identified with the
expressions~\cite{Gutsche:2012bp,Gutsche:2012wb}
\eq\label{Matching1}
M_N(T) = 2 \kappa(T) \sum\limits_\tau\, c_\tau^N\, \sqrt{\tau - 1}
\,, \quad\quad  
M_{\cal R}(T) &=& 2 \kappa(T) 
\sum\limits_\tau\, c_\tau^{\cal R}\, \sqrt{\tau}\,. 
\en
As we mentioned before, the set of mixing parameters $c_\tau^{N,{\cal R}}$ 
is constrained by the correct normalization of the kinetic term of the
four-dimensional spinor field and by charge conservation, as (see detail in 
Ref.~\cite{Gutsche:2012bp}): 
\eq\label{Matching2}
\sum\limits_\tau \, c_\tau^{N,{\cal R}} = 1\,. 
\en 
Finally, the nucleon and Roper masses are expressed in terms of $y$ and
$\sigma(T)$ as
\eq
M_N(T) = M_N \, (1 + y (\sigma(T) - 1))\,,
\quad
M_{\cal R}(T) = M_{\cal R} \, (1 + y (\sigma(T) - 1))\,.
\en 
The baryon form factors are calculated analytically
using the bulk profiles of fermion fields
and the bulk-to-boundary propagator $V(Q,z)$ of 
the vector field (see exact expressions in the next section). 
The calculation technique was described in detail
in Refs.~\cite{Gutsche:2012bp,Gutsche:2012wb,Gutsche:2017lyu}.

\section{Electromagnetic form factors of nucleon, 
Roper and Roper-nucleon transitions}

The electromagnetic form factors of the nucleon and 
Roper-nucleon transitions at finite temperature 
are defined by the following matrix elements, 
due to Lorentz and gauge invariance~\cite{Gutsche:2017lyu},     
\eq\label{matrix_elements}
\vspace*{-.5cm} N \to N: 
M^\mu(p_1\lambda_1,p_2\lambda_2,T) &=& \bar u_N(p_2\lambda_2,T)
\biggl[ \gamma^\mu \, F_1^N(-q^2,T) - 
i \sigma^{\mu\nu} \frac{q_\nu}{2M_N(T)} \, F_2^N(-q^2,T) \, \biggr]
u_N(p_1\lambda_1,T)\,,\\
\vspace*{-.5cm} {\cal R} \to N: 
M^\mu(p_1\lambda_1,p_2\lambda_2,T) &=& \bar u_N(p_2\lambda_2,T)
\biggl[ \gamma^\mu_\perp \, F_1^{{\cal R}N}(-q^2,T) 
-i \sigma^{\mu\nu} \frac{q_\nu}{M_+(T)} \, F_2^{{\cal R}N}(-q^2,T) 
\, \biggr] u_{\cal R}(p_1\lambda_1,T)\,,
\en
where $u_N(p\lambda,T)$ and $u_{\cal R}(p\lambda,T)$ are the 
usual spin-$\frac{1}{2}$ Dirac spinors. They depend on temperature via the masses 
of the nucleon $M_N(T)$ and Roper $M_{\cal R}(T)$, respectively, and 
obey the free Dirac equations of motion 
\eq 
\Big[ \not\! p - M_{N/{\cal R}}(T) \Big] \, u_{N/{\cal R}}(p\lambda,T) = 0\,. 
\en  
In the last set of equations we used
$M_\pm(T) = M_{\cal R}(T) \pm M_N(T)$, $\gamma^\mu_\perp = \gamma^\mu 
- q^\mu \not\! q/q^2$\,, 
$q = p_1 - p_2$, and $\lambda_1$, $\lambda_2$, and $\lambda$ are 
the helicities of the initial, final baryon and photon, which obey the relation 
$\lambda_1 = \lambda_2 - \lambda$. 
 
We recall the definitions of the nucleon Sachs form factors
$G_{E/M}^N(Q^2,T)$, $Q^2 = -q^2$ 
and the electromagnetic radii $\la r^2_{E/M} \ra^N$
in terms of the Dirac $F_1^N(Q^2,T)$ and Pauli $F_2^N(Q^2,T)$ form factors
\eq
G_E^N(Q^2,T) &=& F_1^N(Q^2,T) - \frac{Q^2}{4M_N^2(T)} F_2^N(Q^2,T)\,,
\nonumber\\[2mm] 
G_M^N(Q^2,T) &=& F_1^N(Q^2,T) + F_2^N(Q^2,T)\,, \nonumber\\[2mm]
\la r^2_E(T) \ra^N &=& - 6 \, \frac{dG_E^N(Q^2,T)}{dQ^2}\bigg|_{Q^2 = 0} \,,
\nonumber\\[2mm]
\la r^2_M(T) \ra^N &=&  - \frac{6}{G_M^N(0)} \,
\frac{dG_M^N(Q^2,T)}{dQ^2}\bigg|_{Q^2 = 0}  \,,
\en
where $G_M^N(0, T) \equiv G_M^N(0) \equiv \mu_N$ is the nucleon magnetic moment. 
Note that at $Q^2=0$ the $T$-dependence of all form factors vanishes, while the 
electromagnetic radii are $T$-dependent. As will be seen from our analysis, the 
$Q^2$ behavior of the form factors gets a suppression with increasing temperature.

Now we introduce the helicity amplitudes
$H_{\lambda_2\lambda}$, which in turn can be
related to the invariant form factors $F_i^{{\cal R}N}$ (see details in
Refs.~\cite{Kadeer:2005aq,Faessler:2009xn,Branz:2010pq,Gutsche:2017wag}.
The pertinent relation is 
\eq
H_{\lambda_2\lambda}(T) = M_\mu(p_1\lambda_1,p_2\lambda_2,T)
\, \epsilon^{\ast \, \mu}(q\lambda,T) \,,
\en
where $\epsilon^{\ast \,\mu}(q\lambda,T)$ 
is the polarization vector of the outgoing photon. 
A straightforward calculation 
gives~\cite{Kadeer:2005aq,Faessler:2009xn,Branz:2010pq,Gutsche:2017wag} 
\eq
H_{\pm\frac{1}{2}0}(T) = \sqrt{\frac{Q_-(T)}{Q^2}} \,
\left(
F_1^{{\cal R}N} M_+(T)  - F_2^{{\cal R}N} \frac{Q^2}{M_+(T)} \right) \,, 
\quad \
H_{\pm\frac{1}{2}\pm 1} = - \sqrt{2 Q_-(T)} \,
\left( F_1^{{\cal R}N} + F_2^{{\cal R}N}  \right) \,, 
\en
where $Q_\pm(T) = M_\pm^2(T) + Q^2$. 
In the case of the Roper-nucleon transition there is also 
the set of helicity amplitudes $(A_{1/2}, S_{1/2})$,
related to the set $(H_{\frac{1}{2}0},H_{\frac{1}{2}1})$
by~
\eq
A_{1/2}(T) = - b(T) \, H_{\frac{1}{2}1}(T)\,, \quad
S_{1/2}(T) =   b(T) \, \frac{|{\bf p}(T)|}{\sqrt{Q^2}} \, H_{\frac{1}{2}0}(T)\,,
\en
where
\eq
|{\bf p}(T)| = \frac{\sqrt{Q_+(T) Q_-(T)}}{2M_{\cal R}(T)}\,, \quad  
b(T) = \sqrt{\frac{\pi\alpha}{M_+(T) M_-(T) M_N(T) }}
\en
and $\alpha = 1/137.036$ is the fine-structure constant. 

Expressions for the electromagnetic form factors of the nucleon 
and the Roper-nucleon transition at zero temperature 
are given in Ref.~\cite{Gutsche:2017lyu}. Their temperature dependence 
is generated by the substitution of the dilaton 
parameter at zero temperature with the one at finite $T$: 
$\kappa^2 \to \kappa^2(T)$. 
The parameters which will be used in the numerical evaluations
have been fixed previously and can be found in Ref.~\cite{Gutsche:2017lyu}.  

\begin{figure}
\begin{center}
\epsfig{figure=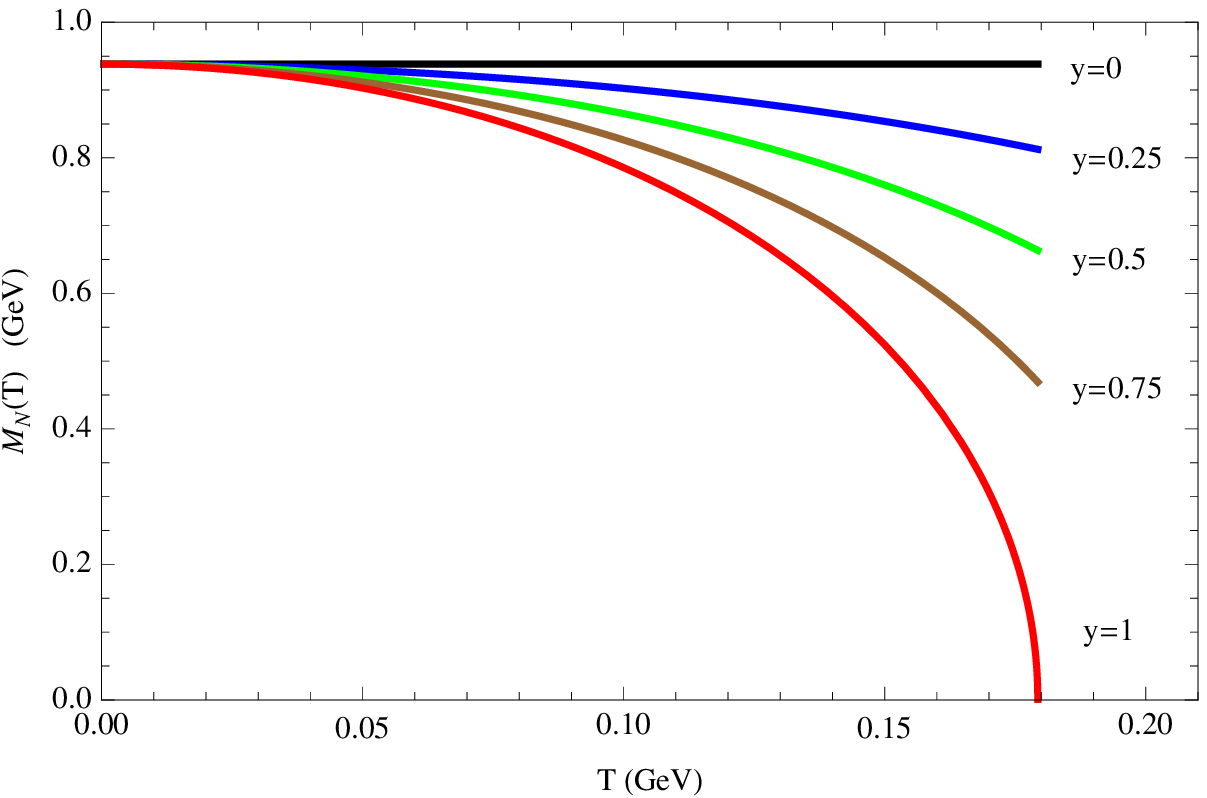,scale=.6}
\epsfig{figure=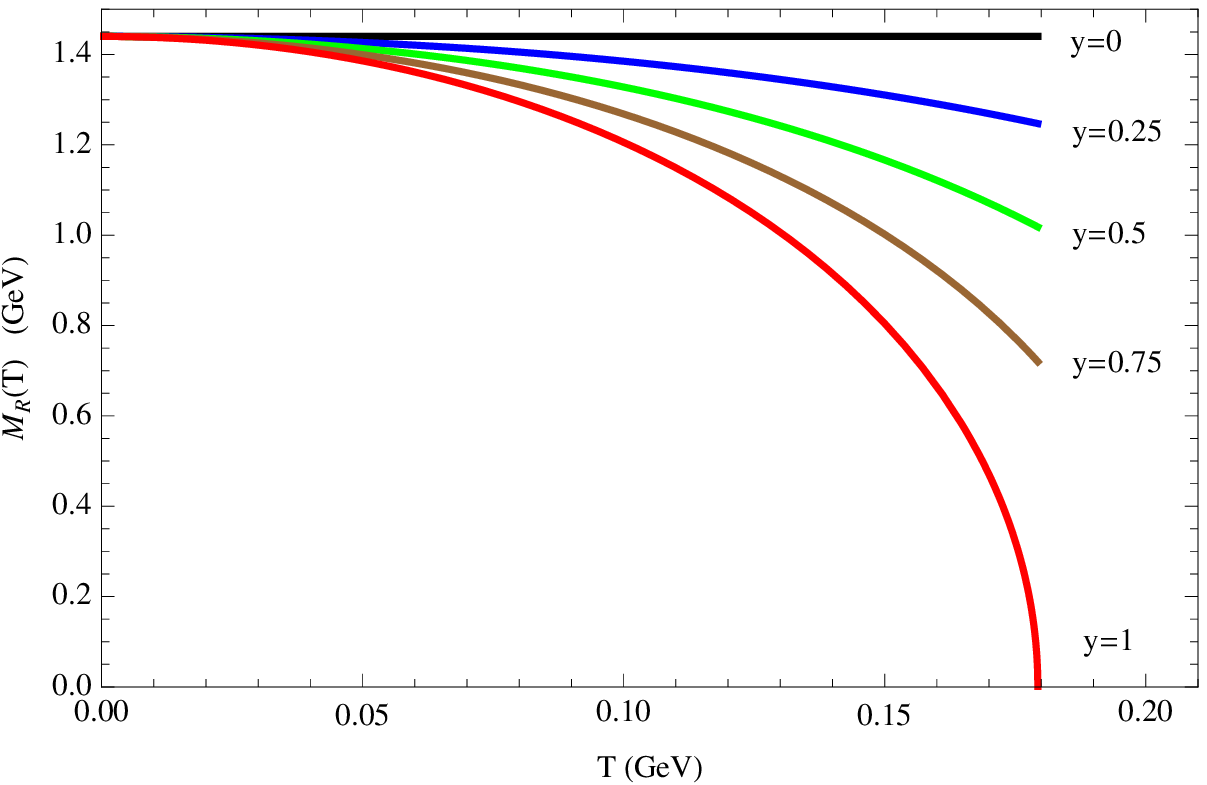,scale=.6}
\end{center}
\noindent
\caption{Temperature dependence of nucleon and Roper masses 
up to $T_c = 230$ MeV.
\label{fig1}}

\vspace*{.5cm}
\begin{center}
\epsfig{figure=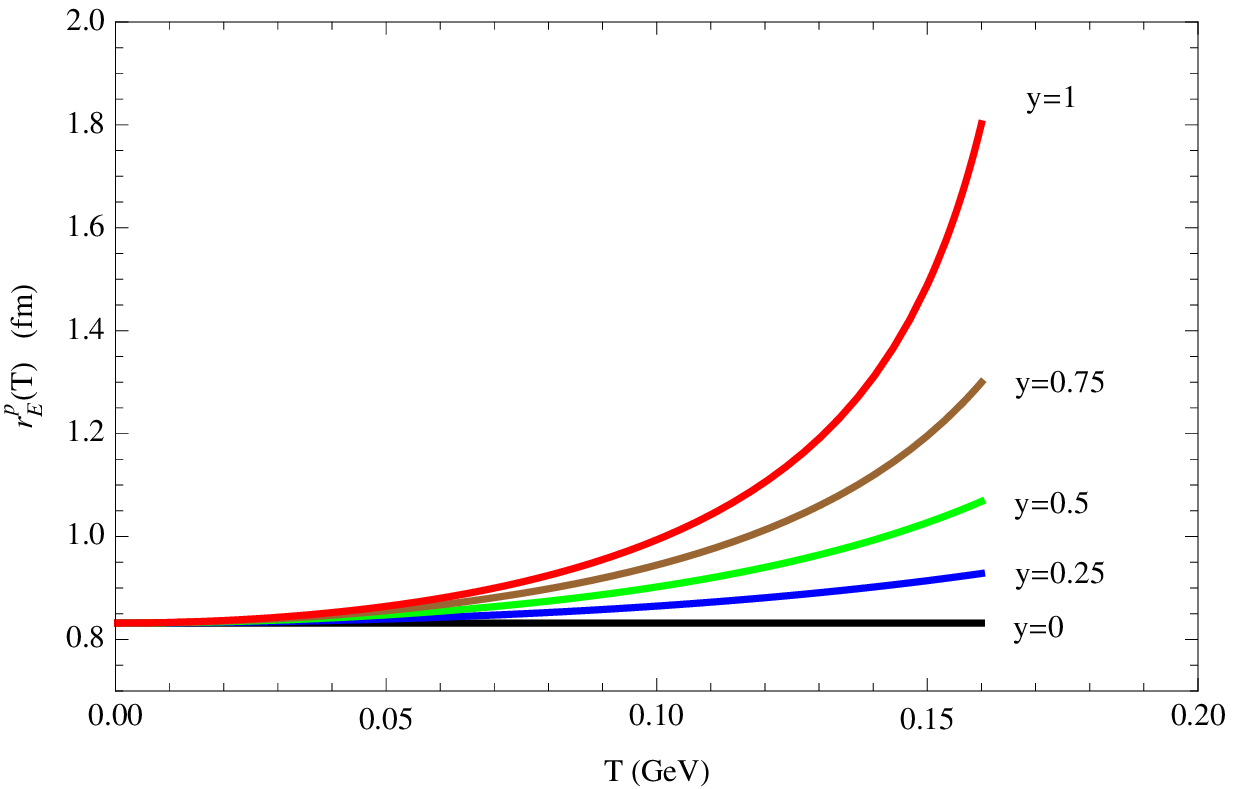,scale=.6}
\epsfig{figure=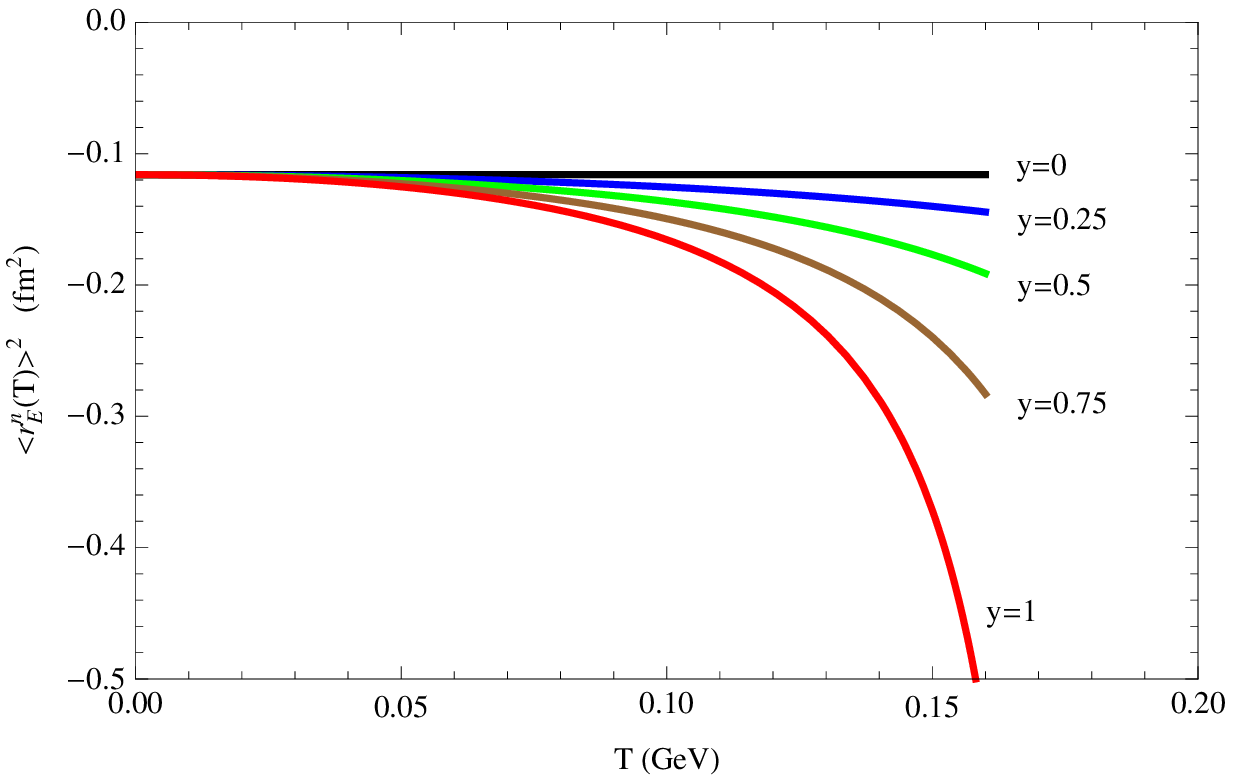,scale=.6}
\end{center}
\noindent
\caption{Temperature dependence of nucleon
charge radii $r_E^p$ and $\la r^2_E \ra^n$ 
up to $T = 200$ MeV. 
\label{fig2}}

\vspace*{.5cm}
\begin{center}
\epsfig{figure=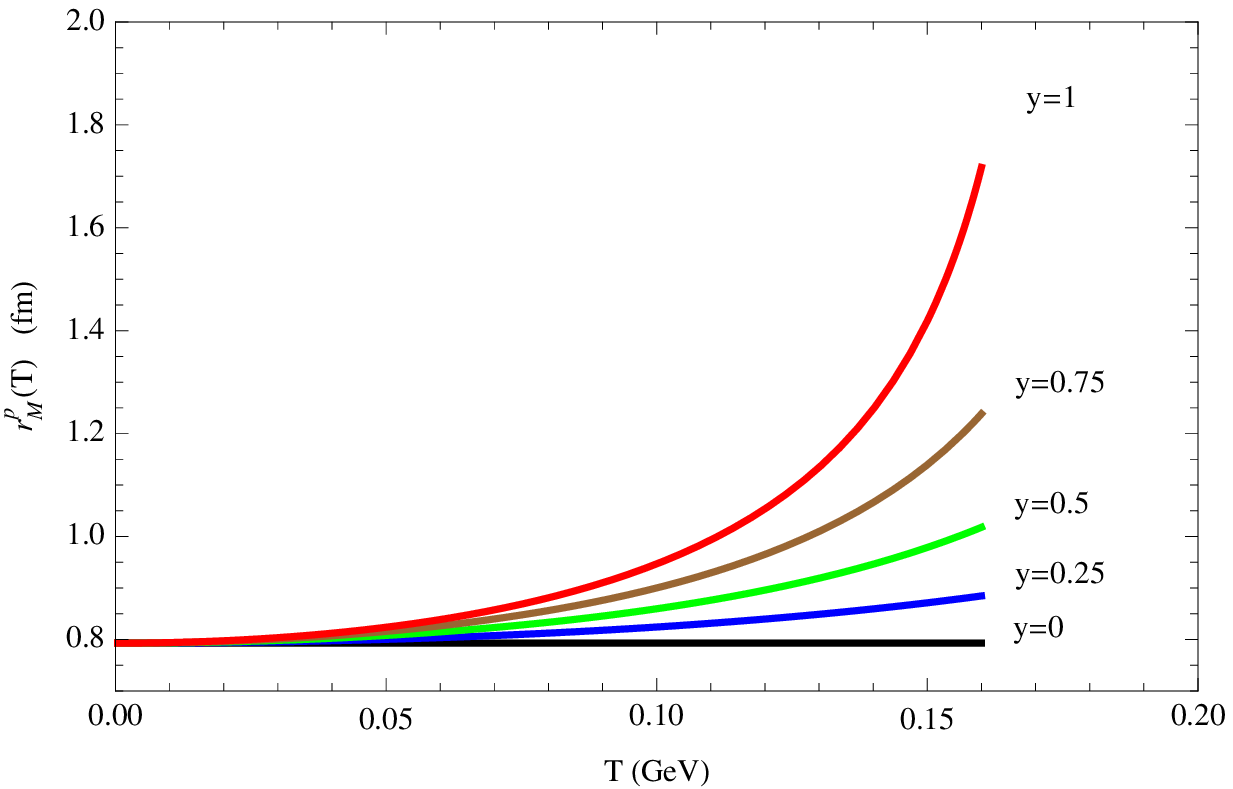,scale=.6}
\epsfig{figure=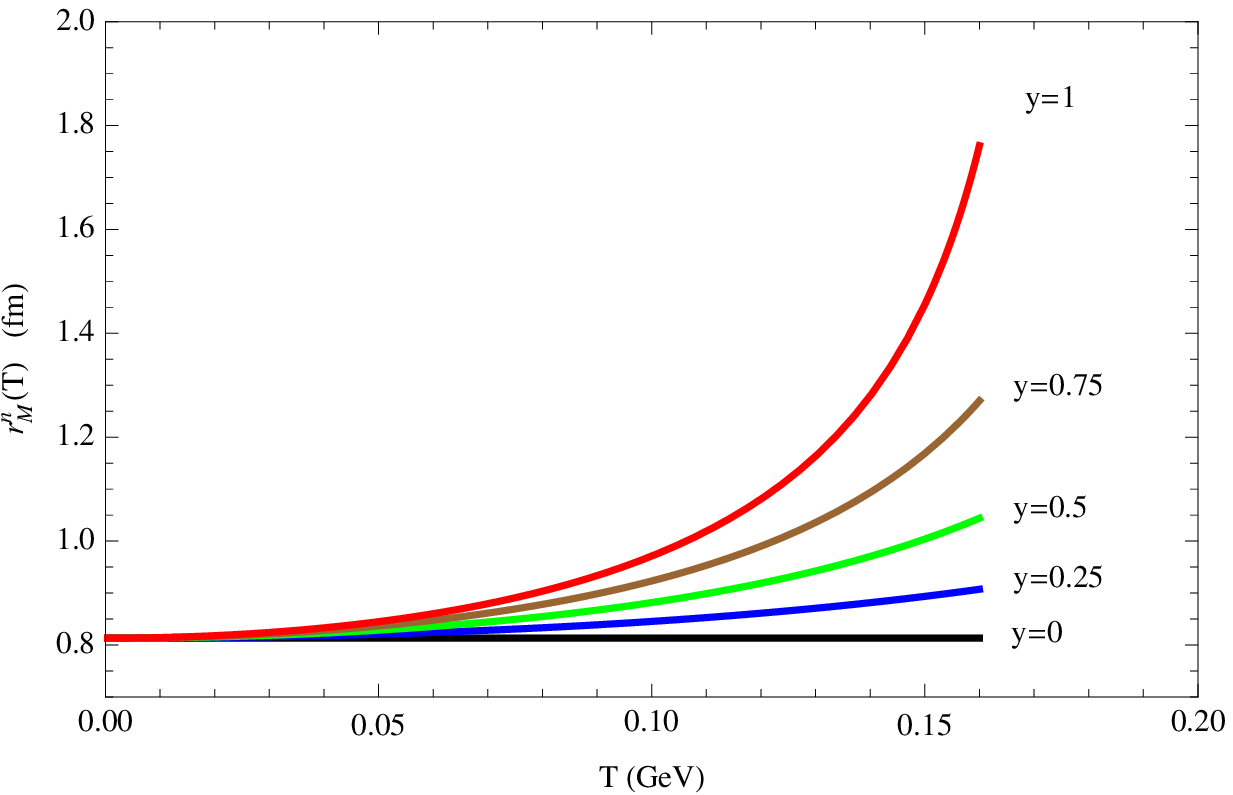,scale=.6}
\end{center}
\noindent
\caption{Temperature dependence of nucleon
magnetic radii $r_M^p$ and $r_M^n$ up to $T = 200$ MeV.
\label{fig3}}
\end{figure}

\begin{figure}
\begin{center}
\epsfig{figure=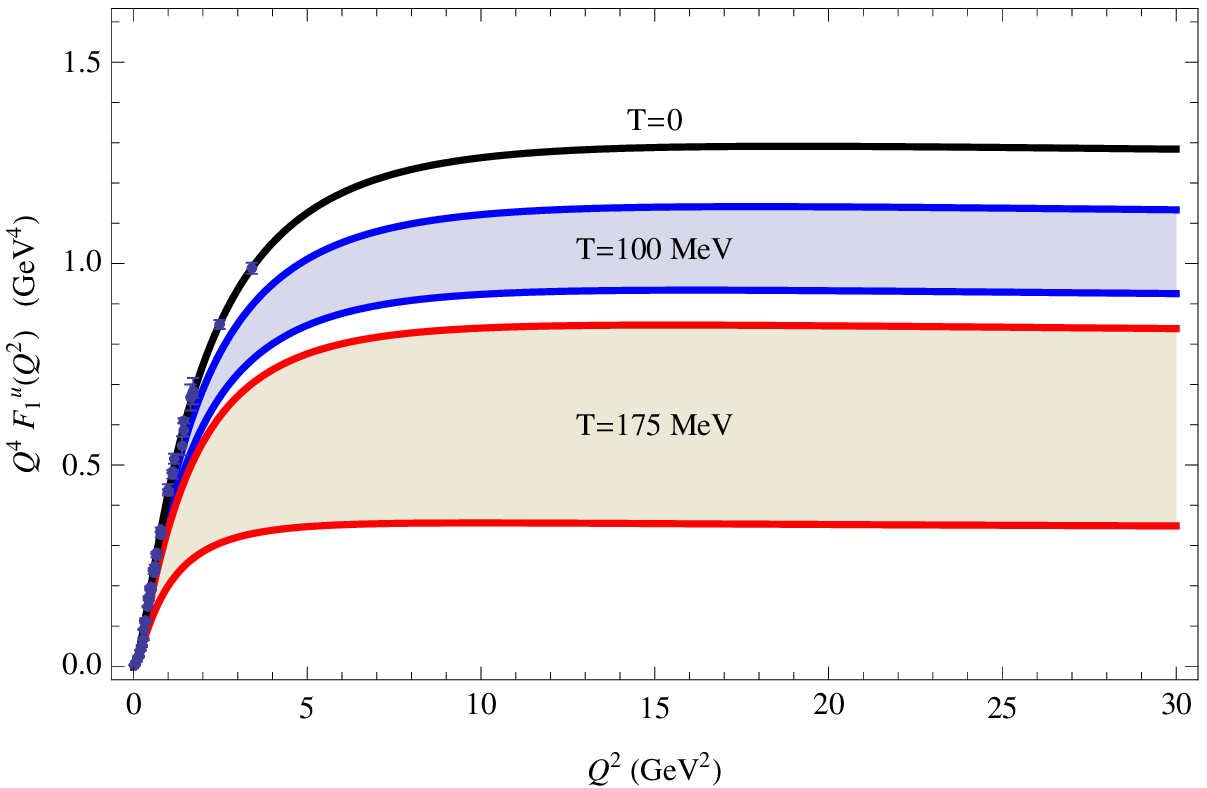,scale=.6}
\epsfig{figure=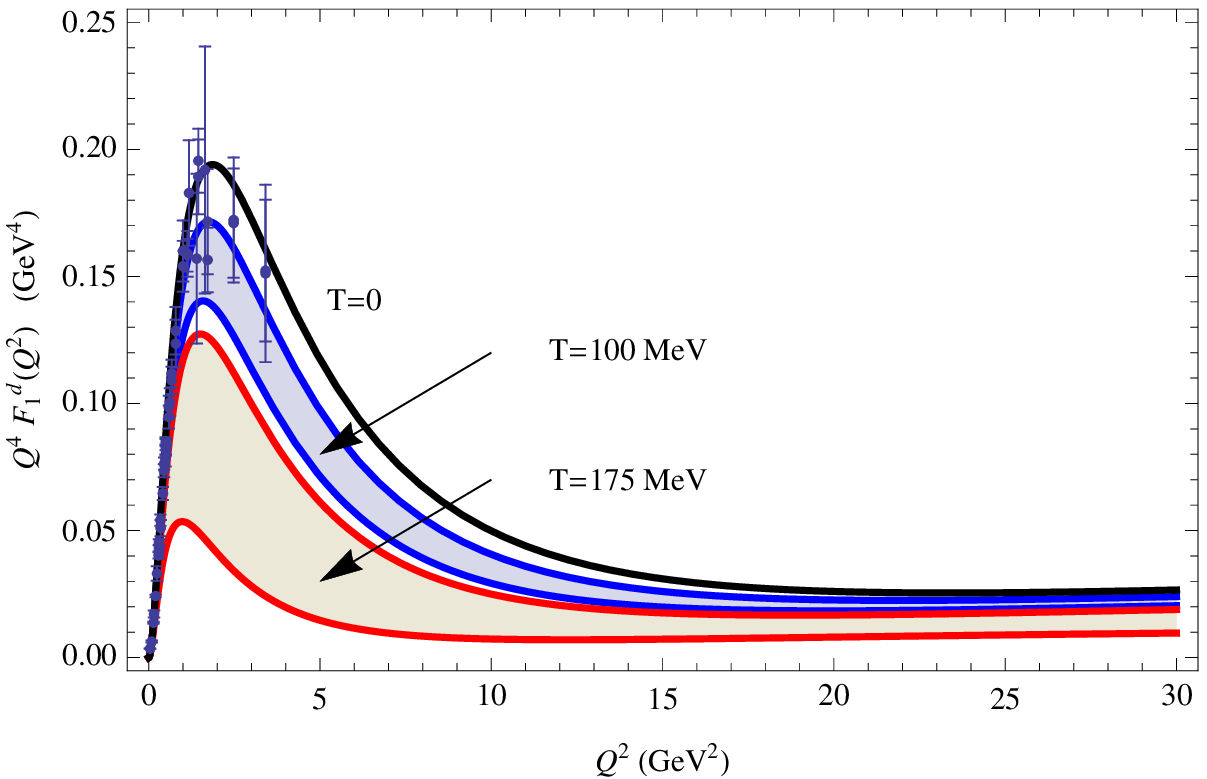,scale=.6}
\end{center}
\noindent
\caption{Temperature dependence of Dirac $u$ and $d$ quark
form factors multiplied by $Q^4$.
\label{fig4}}

\vspace*{.5cm}
\begin{center}
\epsfig{figure=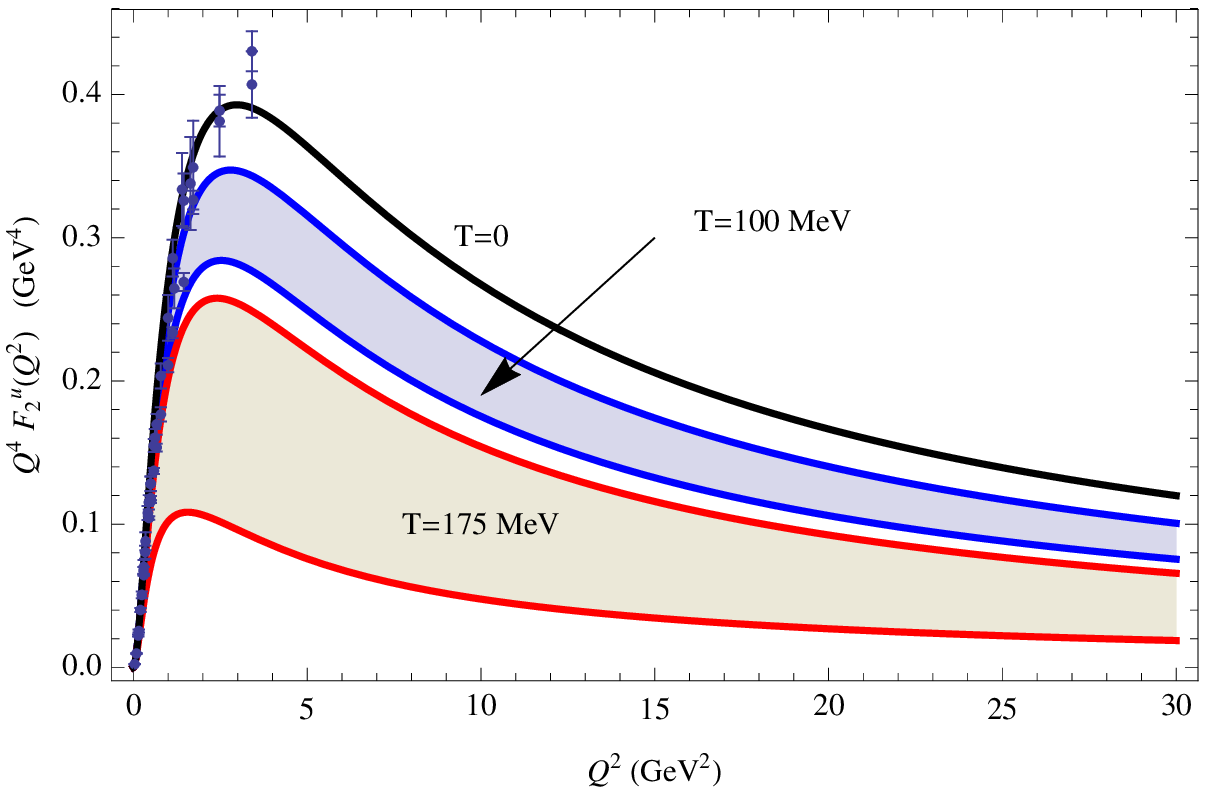,scale=.6}
\epsfig{figure=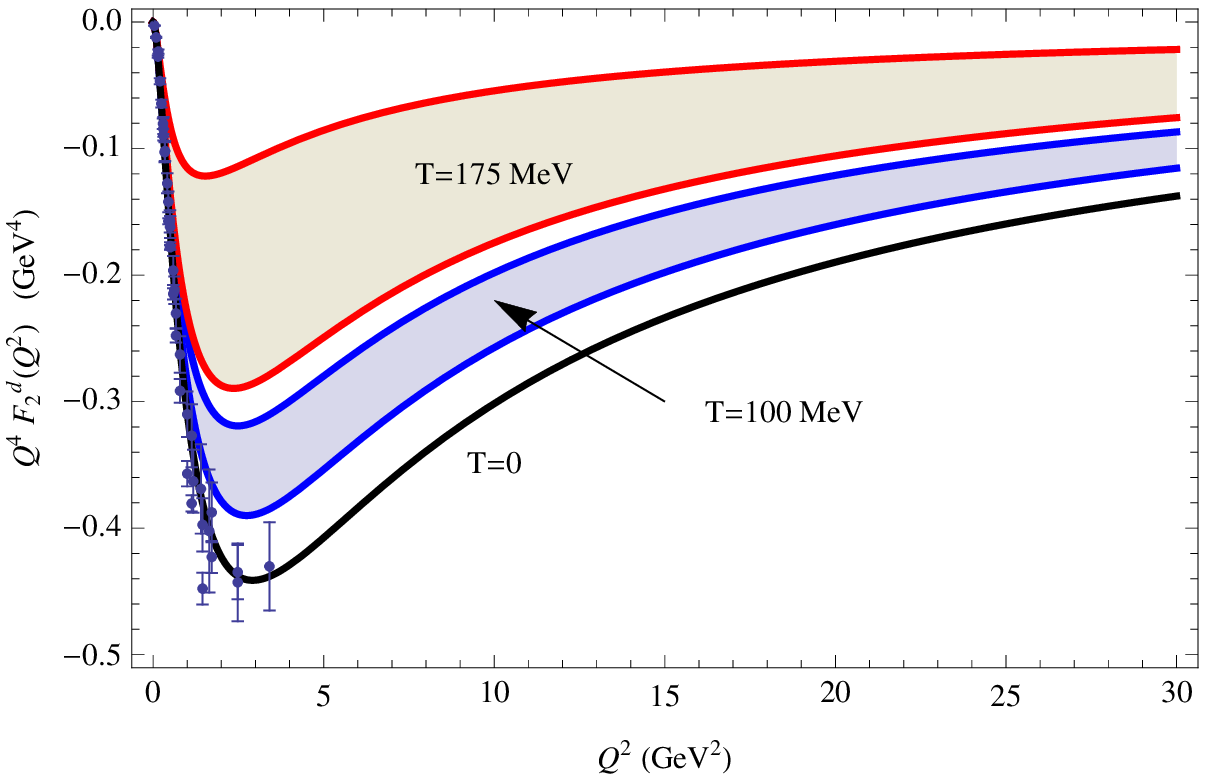,scale=.6}
\end{center}
\noindent
\caption{Temperature dependence of Pauli $u$ and $d$ quark
form factors multiplied by $Q^4$.
\label{fig5}}

\vspace*{.5cm}
\begin{center}
\epsfig{figure=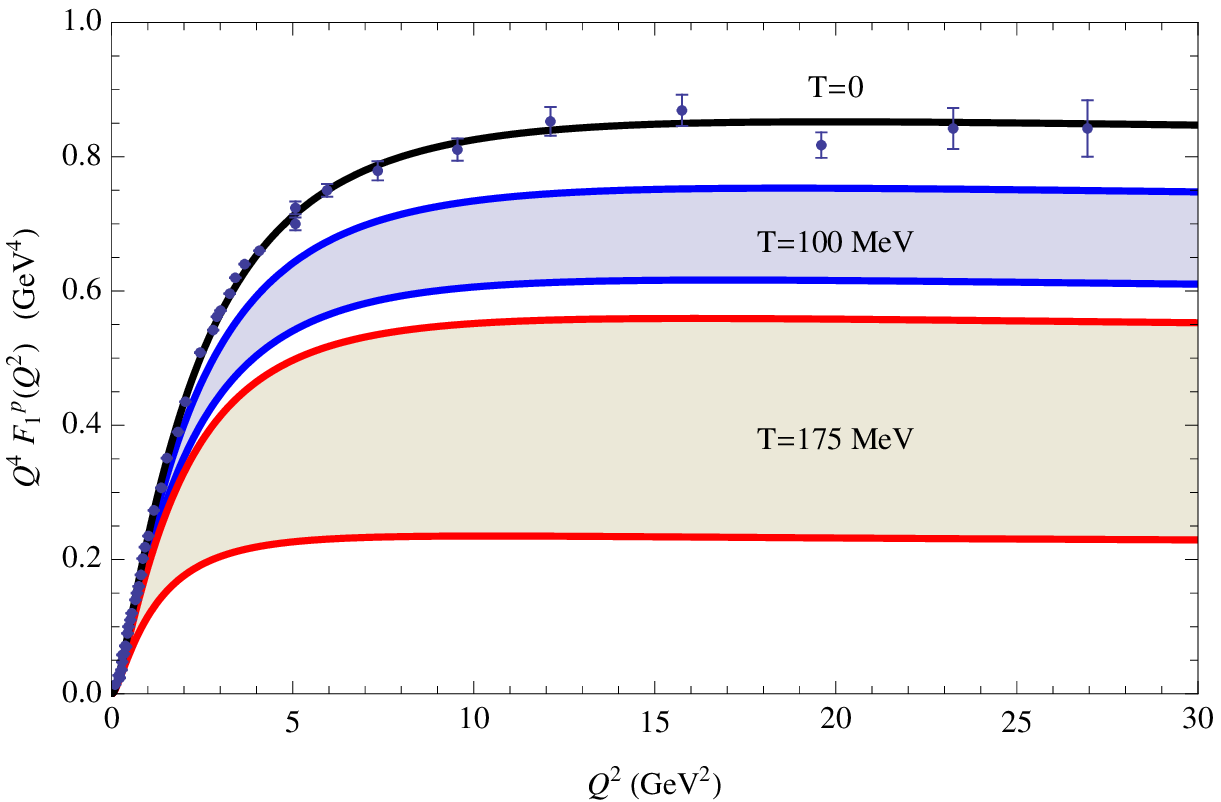,scale=.6}
\epsfig{figure=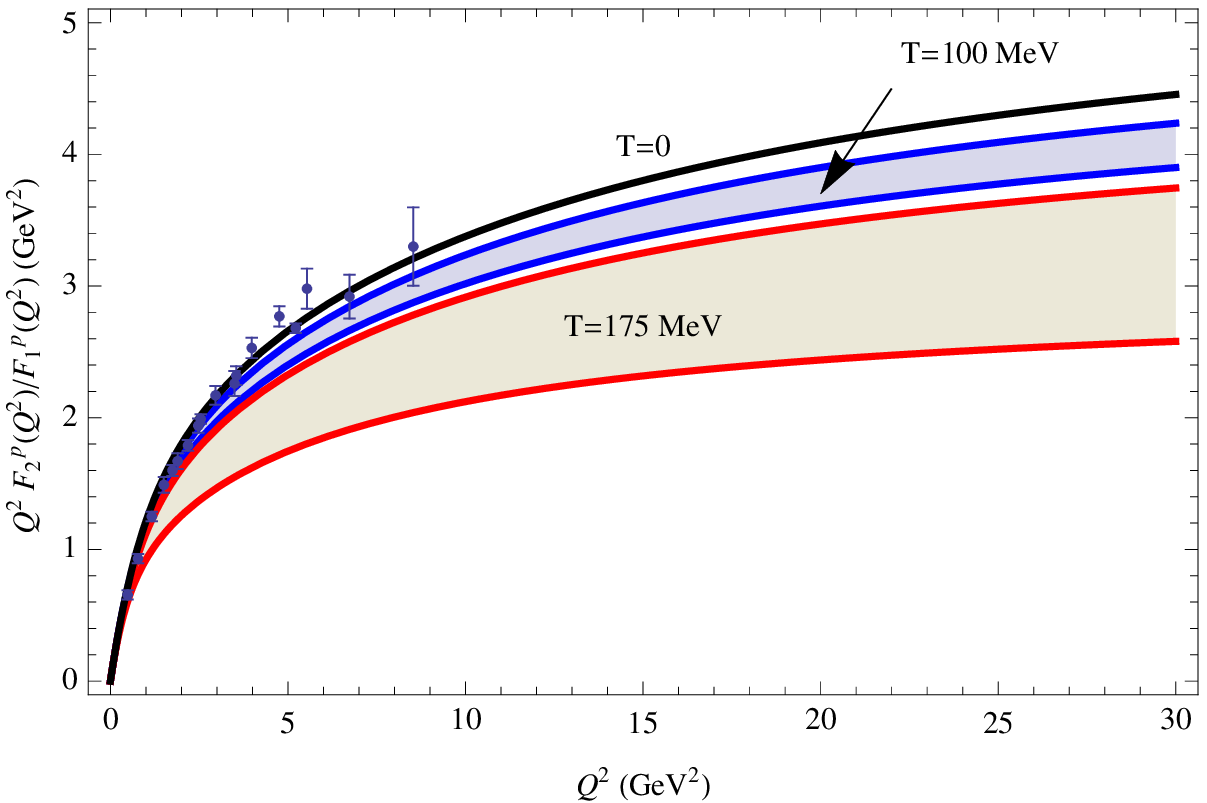,scale=.6}
\end{center}
\noindent
\caption{Temperature dependence of Dirac proton form factor 
multiplied by $Q^4$ and of the ratio $Q^2 F_2^p(Q^2)/F_1^p(Q^2)$.
\label{fig6}}
\end{figure}

\begin{figure}
\begin{center}
\epsfig{figure=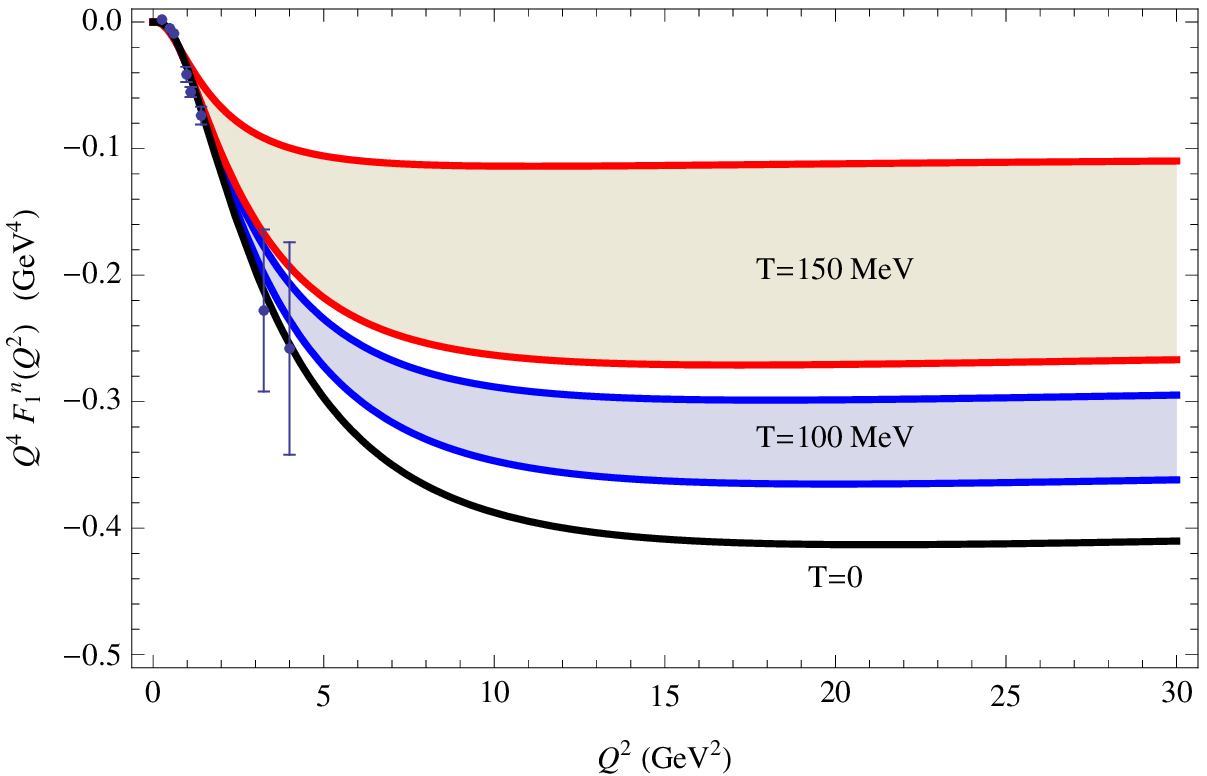,scale=.6}
\epsfig{figure=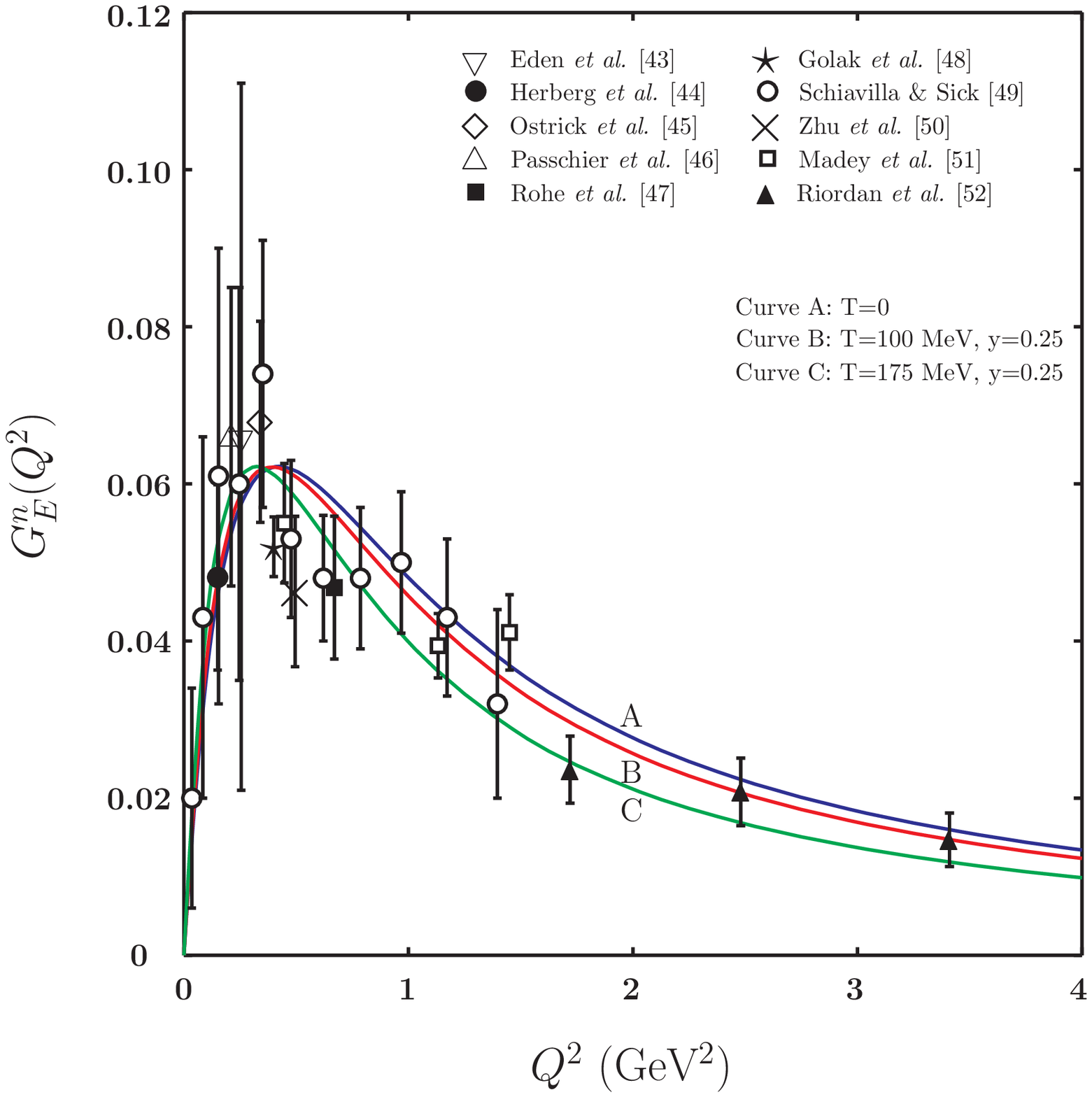,scale=.38}
\end{center}
\noindent
\caption{Temperature dependence of the Dirac neutron
form factor multiplied by $Q^4$ and the charge neutron 
form factor $G_E^n(Q^2)$ in comparison with data.
\label{fig7}}

\vspace*{.5cm}
\begin{center}
\epsfig{figure=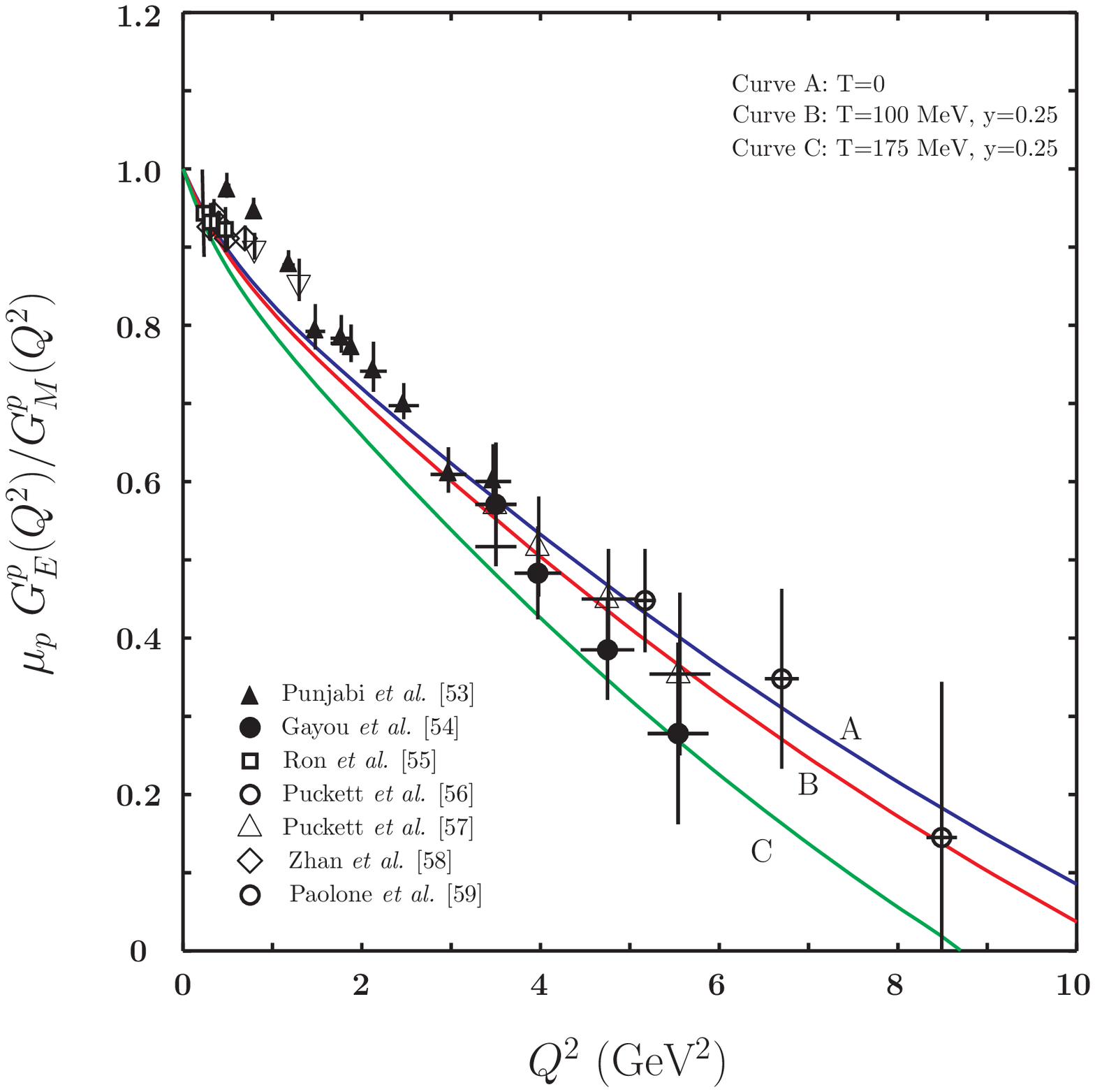,scale=.38}
\epsfig{figure=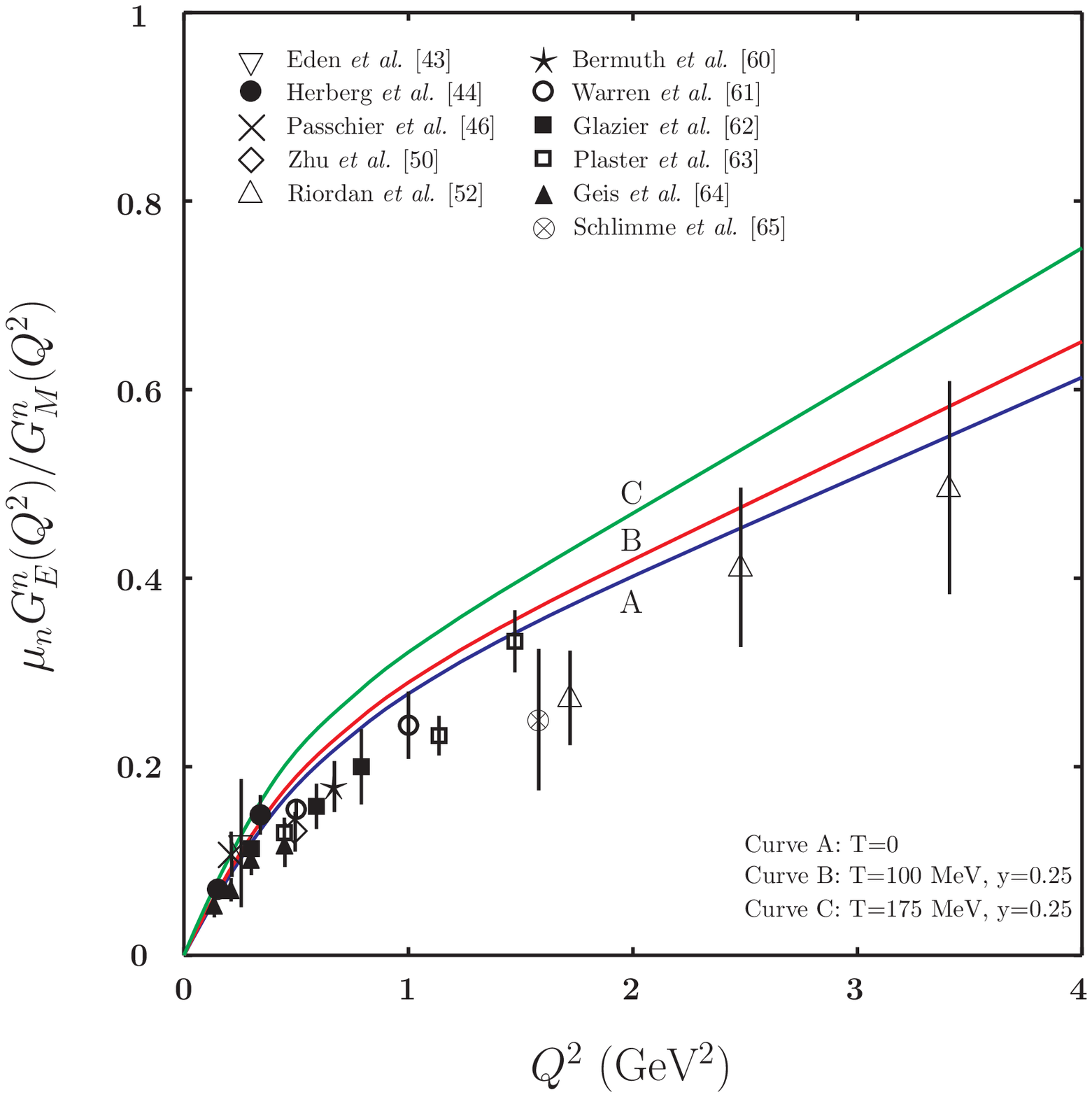,scale=.38}
\end{center}
\noindent
\caption{Temperature dependence of the ratios $\mu_p G_E^p(Q^2)/G_M^p(Q^2)$
and $\mu_n G_E^n(Q^2)/G_M^n(Q^2)$ in comparison with data.
\label{fig8}}
\end{figure}

\begin{figure}
\begin{center}
\epsfig{figure=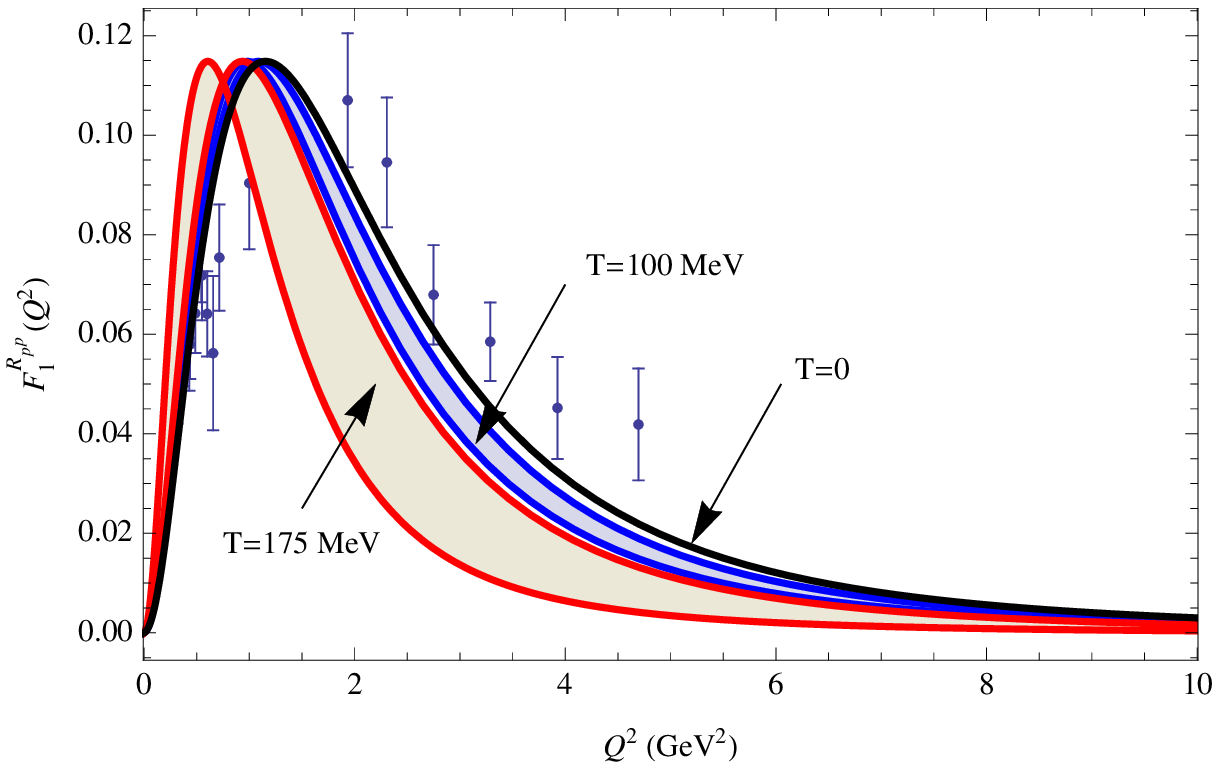,scale=.625}
\epsfig{figure=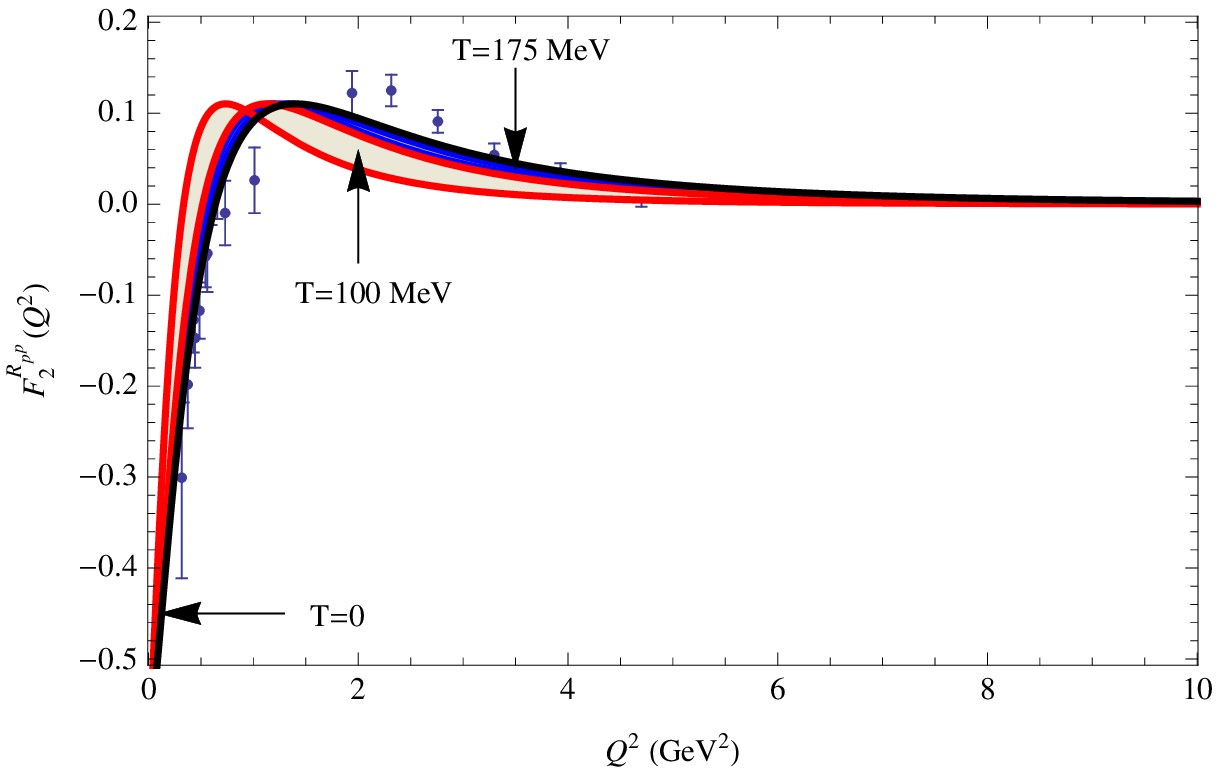,scale=.625}
\end{center}
\noindent
\caption{Temperature dependence of the Roper-nucleon transition form factors
$F_1^{{\cal R}_p p}(Q^2)$ and $F_2^{{\cal R}_p p}(Q^2)$ up to 10 GeV$^2$.
\label{fig9}}

\vspace*{.5cm}
\begin{center}
\epsfig{figure=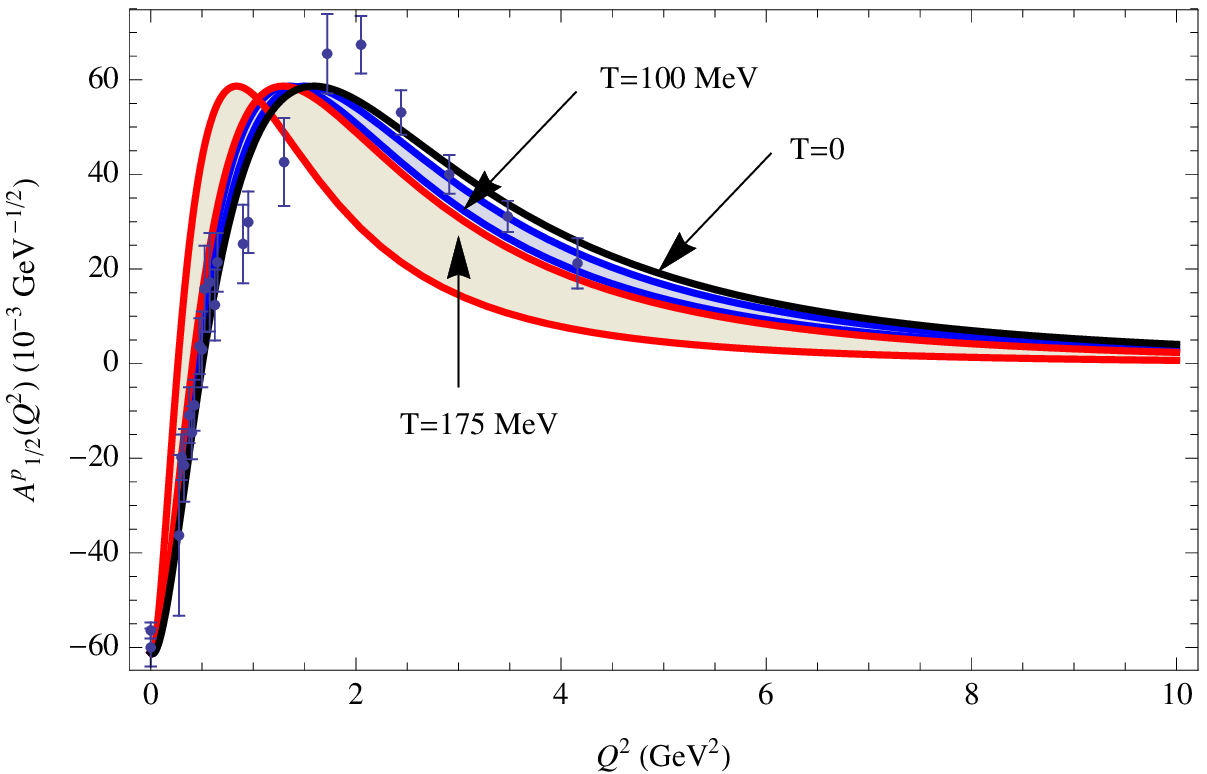,scale=.625}
\epsfig{figure=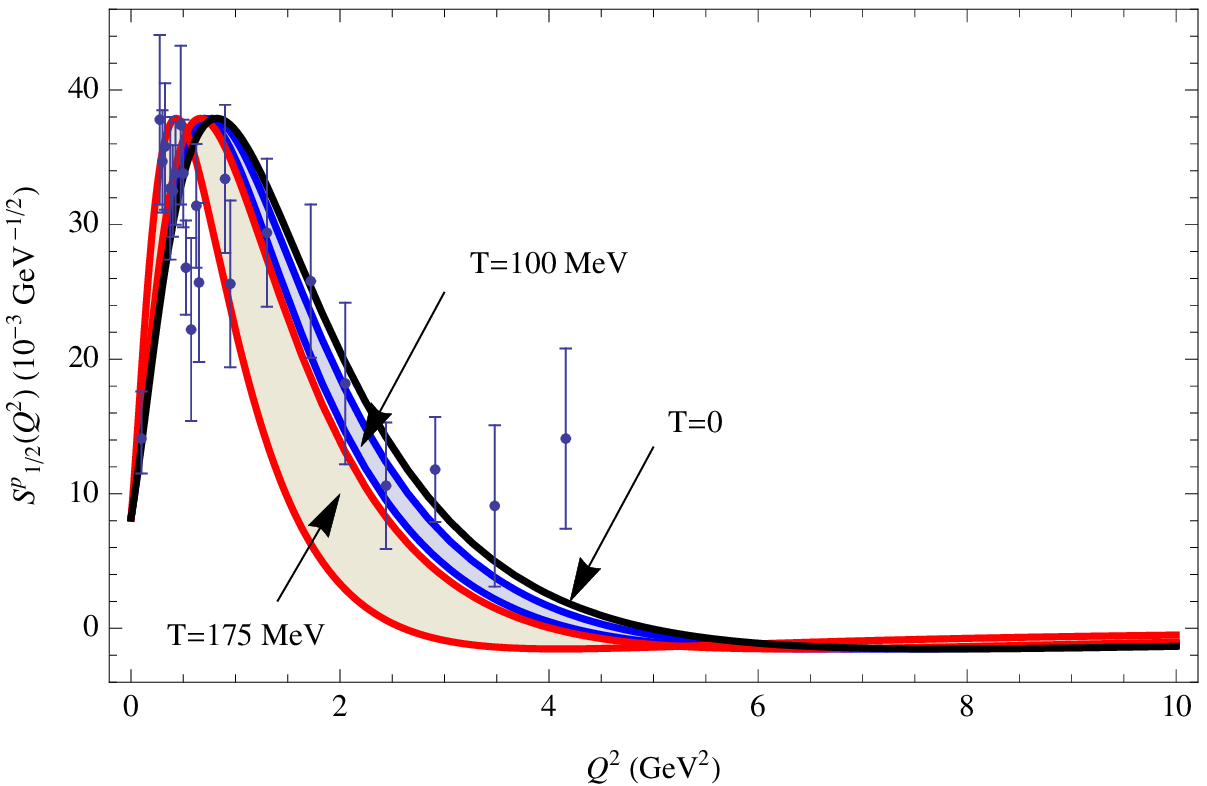,scale=.625}
\end{center}
\noindent
\caption{Temperature dependence of the helicity amplitudes $A_{1/2}^p(Q^2)$
and $S_{1/2}^p(Q^2)$ up to 10 GeV$^2$.
\label{fig10}}
\end{figure}

Our results for the magnetic moments and
Roper-nucleon transition helicity amplitudes at $q^2 = 0$ do not
depend on temperature, while the electromagnetic radii have a 
temperature dependence. The respective results for magnetic moments, charge radii 
at $T=0$ and the Roper-nucleon helicity transition amplitudes at $q^2 = 0$
are summarized in Table I. In Figs.~\ref{fig1}-\ref{fig3} we present our results
for the $T$-dependence from 0 to $T_c=179$ MeV of the nucleon and Roper masses,
and for electromagnetic radii, taking into account the variation of the parameter $y$
in the interval $0 \le y \le 1$. Increasing $y$
and temperature leads to a decrease of the nucleon and Roper mass, while the 
electromagnetic radii of the nucleon are enhanced with increasing $y$ and $T$.
A similar analysis can be done for the form factors and helicity amplitudes.
Our results for the $T$-dependence of quark and nucleon electromagnetic form factors
are shown in Figs.~\ref{fig4}-\ref{fig10}. We compare our results at $T=0$ with
data presented in~\cite{PDG18}-\cite{Stajner:2017fmh}. 
For the electromagnetic form factors we consider three typical values for 
the temperature, $T=0, 100$, and $175$ MeV and vary the parameter $y$ from 0.2
to 0.5. In case of the ratios of the Sachs nucleon form factors, we only considered
a specific value of $y=0.25$. 
The shaded regions in Figs.~\ref{fig4}-\ref{fig7}, \ref{fig9}, and \ref{fig10} 
correspond to the variation of parameter $y$. 
In particular, in Fig.~\ref{fig4} and~\ref{fig5} we present our results for the
Dirac and Pauli $u$ (left panel) and $d$ (right panel) quark form factors.
Here the data were taken from Refs.~\cite{Cates:2011pz,Diehl:2013xca}. 
In Fig.~\ref{fig6} we display the Dirac proton form factor multiplied by $Q^4$
(left panel) and the ratio $Q^2 F_2^p(Q^2)/F_1^p(Q^2)$ (right panel).
In Fig.~\ref{fig7} we show the results for the Dirac neutron form factor multiplied
by $Q^4$ (left panel) and the charge neutron Sachs form factor (right panel).
A detailed comparison of ratios of the nucleon Sachs form factors is shown in 
Fig.~\ref{fig8}, where we use the dipole function $G_D(Q^2)$ with 
$\Lambda^2 = 0.71$ GeV$^2$.
As mentioned before, the $T$-dependence of all form factors vanishes at $Q^2=0$, 
while the electromagnetic radii have an explicit $T$-dependence. 

As it is obvious from our analysis that an increase of temperature leads to 
a suppression of the $Q^2$ behavior of the form factors.
The divergent behavior of the radii and the suppression of
the momentum dependence of form factors is in full agreement
with a previous analysis done in the context of QCD sum 
rules~\cite{Dominguez:1994np} signaling quark deconfinement.

Our predictions for the Roper-nucleon transition form factors and helicity 
amplitudes are shown in Figs.~\ref{fig9} and~\ref{fig10}. At $T=0$ we also 
compare with experimental data of the CLAS 
(JLab)~\cite{Aznauryan:2009mx,Aznauryan:2012ec,Mokeev:2015lda} 
and A1 (MAMI)~\cite{Stajner:2017fmh} Collaborations. 

\begin{table}[ht]
\begin{center}
\caption{Electromagnetic properties of nucleons and Roper}

\vspace*{.1cm}

\def\arraystretch{1.25}
    \begin{tabular}{|c|c|c|}
      \hline
Quantity & Our results & Data~\cite{PDG18}                  \\
\hline
$\mu_p$ (in n.m.)          &  2.793       &  2.793          \\
\hline
$\mu_n$ (in n.m.)          & -1.913       & -1.913          \\
\hline
$r_E^p$ (fm)     &  0.832 &  0.84087 $\pm$ 0.00039 \\
                 &        &  0.8751  $\pm$ 0.0061  \\
\hline
$\la r^2_E \ra^n$ (fm$^2$) & -0.116 & -0.1161 $\pm$ 0.0022  \\
\hline
$r_M^p$ (fm)     &  0.793 &  0.78  $\pm$ 0.04 \\
\hline
$r_M^n$ (fm)     &  0.813 &  0.864$^{+0.009}_{-0.008}$      \\
\hline
$A_{1/2}^p(0)$ (GeV$^{-1/2}$) & -0.061 & -0.060 $\pm$ 0.004 \\
\hline
$S_{1/2}^p(0)$ (GeV$^{-1/2}$) &  0.008 & $\cdots$ \\
\hline
\end{tabular}
\end{center}
\end{table}

\section{Summary}

The present manuscript is based on a soft-wall AdS/QCD approach for the
description of the small temperature dependence of baryon properties.
At small temperatures one can neglect the $T$ dependence of the AdS geometry,
restricting the AdS-Schwarzchild geometry to the AdS Poincar\'e metric.
We study a possible dependence of the holographic potential on the
quark condensate, introducing the mixing coefficient $y$. It is know that
the temperature behavior of the quark condensate dominates the temperature
dependence of gluon condensates of dimension-2 and 4 in the low $T$ region
(at least show the critical temperature of restoration of chiral symmetry)
and starts at order ${\cal T}^2$. Therefore, it is interesting to study
such temperature behavior, because it could be a signal of the contribution
of a quark condensate to baryon properties. As an application we consider
the temperature dependence of the masses and the electromagnetic properties
of the nucleon and the Roper resonance (treated as the first radially
excited state of the nucleon).

\section*{Acknowledgments}

The authors thank Stan Brodsky and Guy T\'eramond for useful discussions.
This work was funded by the Carl Zeiss Foundation under Project
``Kepler Center f\"ur Astro- und Teilchenphysik: Hochsensitive
Nachweistechnik zur Erforschung des unsichtbaren Universums
(Gz: 0653-2.8/581/2)'', by CONICYT (Chile) under
Grants No. 7912010025, 1180232 and PIA/Basal FB0821 and
and by FONDECYT (Chile) under Grant No. 1191103.

\end{document}